\documentclass{iopart}

\usepackage{iopams}

\usepackage[cp866]{inputenc}

\usepackage[russian,english]{babel}
\usepackage{graphicx}
\usepackage{amsfonts}
\usepackage{amssymb}
\usepackage{amsthm}
\usepackage{amscd}
\usepackage{epsfig}

\newcommand{\eqref}[1]{(\ref{#1})}
\newcommand{\rf}[1]{(\ref{#1})}
\def\be{\begin{equation}}
\def\ee{\end{equation}}
\def\bea{\begin{eqnarray}}
\def\eea{\end{eqnarray}}
\def\bal{\begin{align}}
\def\eal{\end{align}}

\newcommand{\ba}[1]{\begin{eqnarray}\label{#1}}
\newcommand{\ea}{\end{eqnarray}}

\newtheorem{Th}{Theorem}
  \newtheorem{Lm}{Lemma}

\def\RR{\mathbb{R}}

\def\CC{\mathbb{C}}

\def\ZZ{\mathbb{Z}}

\newcommand{\cD}{\mathcal{D}}
\newcommand{\cH}{\mathcal{H}}

\newcommand{\cL}{\mathcal{L}}

\newcommand{\spec}{\mathrm{spec}}
\newcommand{\for}{\mathrm{for\ }}

\newcommand{\bfPsi}{\mathbf{\Psi}}
\newcommand{\bK}{\mathbf{K}}

\def\p{\partial}
\def\a{\alpha}

\def\k{\kappa}

\def\T{\Theta}

\def\wt{\widetilde}

\def\erfc{{\rm erfc}\,}
\def\erf{{\rm erf}\,}
\def\exp{{\rm exp}\,}
\def\res{{\rm res}\,}
\newcommand{\Sh}{Schr\"odinger\ }
\renewcommand\Im{{\rm Im}}

\newcommand{\nn}{\nonumber}

\catcode`\@=11
\renewcommand\footnoterule{%
  \kern-3\p@
  \hrule\@width.4\columnwidth
  \kern2.6\p@}
\renewcommand\@makefntext[1]{%
    \parindent 1em\noindent
    \hb@xt@1.8em{\hss$^{\@thefnmark}$)}\hspace{2pt}%
    \footnotesize\rmfamily#1}  
\def\@makefnmark{\hspace{.5pt}\hbox{$^{\@thefnmark}$%
\hspace{-1pt})}} 

\begin{document}
\title[Exact propagators for SUSY partners]{Exact propagators for
SUSY partners}

\author{Andrey M Pupasov$^1$, Boris F Samsonov$^1$ and \\ Uwe G\"unther$^2$}

\address{$^1$ Physics Department, Tomsk State University, 36 Lenin Avenue,
634050 Tomsk, Russia}
\address{$^2$ Research Center Dresden-Rossendorf,  POB 510119,
D-01314 Dresden, Germany}

\eads{ \mailto{pupasov@phys.tsu.ru}, \mailto{samsonov@phys.tsu.ru}
and \mailto{u.guenther@fzd.de}}

\begin{abstract}
Pairs of SUSY partner Hamiltonians are studied which are
interrelated by usual (linear) or polynomial supersymmetry.
Assuming the model of one of the Hamiltonians as exactly solvable
with known propagator, expressions for propagators of
partner models are derived.
The corresponding general results are
applied to ``a particle in a box'', the Harmonic oscillator and a
free particle (i.e. to transparent potentials).
\end{abstract}
{\small submitted to:} {\it J. Phys. A: Math. Theor.}


\section{Introduction}

The space-time evolution of a quantum mechanical system is
governed by its \Sh equation and in its most complete form it is
encoded in the propagator. The propagator defines the probability
amplitude for a particle to move from one point of the space
 to another in a
given time. Similar to propagators in relativistic field theories,
it provides a global picture of the causal structure of a quantum
system which goes beyond the information contained in a single
wave function. Moreover, it plays the essential role in solving
the probability related Cauchy problem of Quantum Mechanics (QM).

The vast literature on QM propagators, summarized e.g. in
\cite{Grosche}, lists mainly explicit expressions of propagators for
stationary \Sh equations in one space dimension which are reducible
to hypergeometric differential equations (DEs) or their confluent
forms. This is in strong contrast to the available methods of
supersymmetric Quantum Mechanics (SUSY QM) \cite{Sukumar1,BS} which
usually lead to much broader classes of exactly solvable \Sh
equations with  solutions in particular expressible in terms of
linear combinations of hypergeometric functions (see e.g.
\cite{SO}). From the structure of the SUSY induced relations between
superpartner Hamiltonians it is clear that via the corresponding \Sh
equations these relations should  extend to relations between the
associated propagators. The main goal of the present paper is to
analyze these relations between SUSY partner propagators and to
reshape them into user friendly general recipes for the construction
of new propagator classes. Our interest in such new exact
propagators is less motivated by their mere existence or their
technical subtleties, but rather it is in their applicability to
concrete physical problems (see e.g. \cite{JPA}) and here especially
to models with well-tailored new properties and to setups which up
to now were not related to SUSY techniques at all (see e.g.
\cite{FisicaE}). In order to derive the corresponding technical
tools we concentrate in this paper on the general approach which
allows to establish the link between the propagators of any two SUSY
partner Hamiltonians.

Our main idea is the following.
For a given Hamiltonian $h_0$,
 whose \Sh equation
as a second order differential equation
 is exactly solvable,
 the general
solutions of the \Sh equation of its SUSY partner Hamiltonian $h_N$
can be explicitly constructed.
Knowing the relations between the
solutions of the \Sh equations for $h_0$ and $h_N$ one can expect to
derive transformation operators relating the associated propagators.
At a first glance, this approach seems to be technically trivial.
But when one recalls that the calculation of a closed expression for
a propagator is usual more difficult than the derivation of a
corresponding wave function, one may expect the problem to be
connected with rather nontrivial technical subtleties. Moreover, if
one wants to establish the explicit link between the propagators one
may imagine that sometimes the problem may even become unsolvable.

Probably the first indication that the problem may have a solution
was given by Jauslin \cite{jauslin} who constructed a general
integral transformation scheme simultaneously for \Sh equations and
for heat equations, but who didn't provide a discussion of
convergency and divergency of the derived expressions.
For the sake of
convergency he applied his technique to the heat-equation-type
Fokker-Planck equation only. In general, this result may be extended
via Wick rotation to propagators for  \Sh equations of a free
particle and a particle moving through transparent potentials. But
the question of convergency and with it the question of solvability
remains to be clarified. Another indication that the problem may be
solvable has been provided by Refs \cite{SSP,SP} where a similar
model has been analyzed at the level of Green functions of
stationary \Sh equations. In the present paper we carefully analyze
the problem along the ideas first announced in \cite{SPLA}.

The material is organized as follows. In the next section
\ref{prelim}, we briefly recall some main facts and notations from
SUSY QM necessary for our subsequent analysis and we sketch the
definition of the propagator. Section \ref{SUSY-1} is devoted to
the basic tool of our approach ---  the interrelation between
propagators for models with partner Hamiltonians which are linked
by first-order SUSY transformations. In Section \ref{SUSY-N} we
generalize these results to polynomial supersymmetry. The general
technique developed in these sections is afterwards used in
section \ref{applic} to derive the propagators for SUSY partner
models of ``a particle in a box'', of the Harmonic oscillator and
of the free particle (i.e. for models with transparent
potentials). Section \ref{conclu} concludes the paper.

\section{Preliminaries\label{prelim}}

\selectlanguage{english}

\vspace{1em}

The subject of our analysis will be a polynomial
generalization\footnote{For a recent review see e.g. \cite{AC}.}
of the simplest two-component system of Witten's non-relativistic
supersymmetric quantum mechanics
 \cite{Witten,Sukumar1,BS} described by the \Sh equation
\begin{equation}\label{supershredinger}
(\rmi I \partial_t-H){\mathbf{\Psi}}(x,t) =0 \qquad x\in (a,b)
\end{equation}
where  $H$ is a diagonal super-Hamiltonian consisting of the two
super-partners $h_0$ and $h_{N}$ as components
\begin{equation}\label{superhamiltonian}
H= \left(
\begin {array}{cc} h_0&0\\ \noalign{\medskip}0&h_N\end{array}
 \right)
 \qquad h_{0,N}=-\partial^2_{x}+V_{0,N}(x)\,.
\end{equation}
The interval $(a,b)$ may be both finite or infinite. For simplicity
we restrict our consideration to a stationary setup with $h_0$ and
$h_{N}$ not explicitly depending on time so that the evolution
equation
 \rf{supershredinger} reduces via standard substitution
${\bfPsi}(x,t)=\bfPsi(x)\mathrm{e}^{-iEt}$ and properly chosen
boundary conditions to the spectral problem
\begin{equation}\label{stsh}
H\bfPsi=E\bfPsi\,.
\end{equation}
The time evolution of the system may be described in terms of a
corresponding propagator.

Furthermore, we assume that the partner Hamiltonians $h_0$ and
$h_{N}$ are intertwined by an $N$th-order differential operator $L$
with the following properties: \\ {\bf 1. intertwining relations}
\be
 Lh_0=h_NL \qquad h_0L^+=L^+h_N \label{intertwiner}
\ee
{\bf 2. factorization rule}
\bea\nonumber
&& L^+L=P_N(h_0) \qquad LL^+=P_N(h_N)\\ \nonumber &&
P_N(x)=(x-\a_0)\ldots(x-\a_{N-1})\,\label{polyn-1}\\  && \Im(\a_i)=0
\qquad \a_i\neq \a_{k\neq i}\qquad  i,k=0,\ldots,N-1\,.
\eea
Here, the adjoint operation is understood in the sense of Laplace
(i.e. as formally adjoint with the property $\p_x^+=-\p_x$,
$(AB)^+=B^+A^+$ and $\rmi^+=-\rmi$) and the roots $\a_i$ of the
polynomial $P_N$ play the role of factorization constants. For
simplicity we assume that the polynomial $P_N$ has only simple
roots. The intertwining relations together with the factorization
rule can be represented in terms of the polynomial super-algebra
\cite{AIS,AC,BS}
\be\label{superalgebra} \fl
Q^2=(Q^+)^2=0\qquad [Q,H]=[Q^+,H]=0\qquad QQ^++Q^+Q=P_N(H)
\ee
with nilpotent super-charges
\[
Q=
\left(
\begin{array}{cc}
0 & 0 \\
L & 0
\end{array}
\right)\qquad
Q^+=
\left(
\begin{array}{cc}
0 & L^+ \\
0 & 0
\end{array}
\right).
\]

Although the component Hamiltonians $h_0$ and $h_N$ enter the
super-Hamiltonian \eqref{superhamiltonian} in an algebraically
symmetric way, we consider $h_0$ as given Hamiltonian with known
spectral properties and $h_N$ as derived Hamiltonian with still
undefined spectrum. More precisely we assume $V_0(x)$ to be
real-valued, continuous and bounded from below%
\footnote{We assume the potential $V_0(x)$ bounded from below and
short ranged, because we restrict our attention here to models with
a finite number of bound states only.  For spectral problems on the
half-line a repulsive singularity at the origin not stronger than
$\ell(\ell+1)x^{-2}$, $\ell=0,1,\ldots$ is possible.} so that the
differential expression $h_0=-\p_x^2+V_0(x)$ defines a
Sturm-Liouville operator which is symmetric with respect to the
usual $\cL^2(a,b)$ inner product. The corresponding functions
$\psi\in\cL^2(a,b)$ are additionally assumed sufficiently
smooth\footnote{As usual, $C^2(a,b)$ denotes the space of twice
continuously differentiable functions.}, e.g. $\psi\in C^2(a,b)$,
over the interval $(a,b)\subseteq \RR$. Moreover, we assume
Dirichlet boundary conditions (BCs) for the bound state
eigenfunctions of $h_0$, i.e a domain $\cD(h_0):=\{\psi:\,\psi\in
\cL^2(a,b)\cap C^2(a,b),\, \psi(a)=\psi(b)=0 \}$  (see, e.g.
\cite{kost}) and the operator $h_0$ itself being at least
essentially self-adjoint (with a closure that we denote by the same
symbol $h_0$). As usual, eigenfunctions which correspond to the
continuous spectrum of $h_0$ are supposed to have an oscillating
asymptotic behavior at spatial infinity. Concentrating on physically
relevant cases we restrict our attention to the following three
types of
setups:\\
{\bf (i):} \ \ The interval $(a,b)$ is finite $|a|,|b|<\infty$ so
that $h_0$ has a  non-degenerate purely
discrete spectrum (see e.g. \cite{Levitan}).\\
{\bf (ii):} \ \ For spectral problems on the half-line
$(a=0,b=\infty)$ we consider so called scattering (or
short-ranged) potentials which decrease at infinity faster than
any finite power of $x$ and have a continuous spectrum filling the
positive semi-axis and a finite number of discrete levels; the
whole spectrum is
non-degenerate.\\
{\bf (iii):} \ \ For spectral problems on the whole real line,
$(a=-\infty,b=\infty)$, we consider confining as well as scattering
potentials. Confining potentials produce purely discrete
non-degenerate spectra (see e.g. \cite{Berezin}), whereas scattering
potentials lead to two-fold degenerate continuous spectra filling
the whole real line and to a finite number of non-degenerate
discrete levels (see e.g. \cite{Levitan}).

Everywhere in the text we choose real-valued solutions of the
differential equation $(h_0-E)\psi=0$. This is always possible since
$V_0(x)$ is supposed to be a real-valued function and we always
restrict ourselves to real values of the parameter $E$.

The intertwiner $L$ is completely described by a set of $N$
transformation functions $u_n(x)$, which may be both
``physical'' and ``unphysical''%
\footnote{By ``physical'' solutions we mean solutions belonging to
the domain $\cD(h_0)=\{\psi:\ \psi\in \cL^2(a,b)\cap C^2(a,b), \
\psi(a)=\psi(b)=0 \}$. All solutions $\psi\not\in \cD(h_0)$
corresponding to a spectral parameter $E$ outside the continuous
spectrum  are interpreted as ``unphysical''. In the present paper,
eigenfunctions corresponding to the continuous spectrum are not
used as transformation functions and, although physically
meaningful, we excluded them from our classification scheme of
``physical'' and ``unphysical'' solutions.} solutions of the
stationary \Sh equation with $h_0$ as Hamiltonian:
\[h_0u_n=\a_nu_n \quad n=0,\ldots,N-1\,.\]
In our case of a polynomial $P_N(x)$ with simple roots (i.e.
$\a_i\neq \a_k$) the action of the intertwiner $L$ on a function $f$
is given by the Crum-Krein formula \cite{CK,Krein}
\be\label{operator}
Lf=\frac{W(u_0,u_1,\ldots,u_{N-1},f)}{W(u_0,u_1,\ldots,u_{N-1})}
\ee
with $W$ denoting the Wronskian
\be
W=W(u_0,u_1,\ldots,u_{N-1})=
\left|\begin{array}{cccc}u_0 & u_1&
\ldots &u_{N-1}\\
u'_0&u'_1&\ldots&u'_{N-1}\\
\ldots&\ldots&\ldots&\ldots\\
u_0^{(N-1)}&u_1^{(N-1)}&\ldots&u_{N-1}^{(N-1)}\\
\end{array}\right|.
\ee
It links the solutions $\phi$ and $\psi$ of the \Sh equations with
$h_N$ and $h_0$ as Hamiltonians by the relation $\phi = L\psi$. In
particular, for $N=1$ (a first-order transformation)
\eqref{operator} reads \be\label{L1}
\phi=Lf=(-u_{0x}/u_0+\p_x)f=\frac{W(u_0,f)}{u_0}
\ee
with $W(u_0)\equiv u_0$. Furthermore, the determinant structure
\rf{operator} of the operator $L$ leads to the immediate implication
that it has a nontrivial kernel space $\mathrm{Ker}L$ spanned by the
set of transformation functions $u_n$:
\[
\mathrm{Ker}L = \mathrm{span}\{u_0,\ldots,u_{N-1} \}\qquad
\mathrm{dim} (\mathrm{Ker}L)=\mathrm{dim} (\mathrm{Ker}L^+)=N\,.
\]
The solutions $v_n$ of the equation $h_Nv_n=\alpha_nv_n$ are
elements of the kernel space of the adjoint operator $L^+$ and can
be obtained as\footnote{See also \rf{Wn-frac} in \ref{ap1}.}
\be\label{hNsol}
v_n= \frac{ W_n(u_0,u_1,\ldots,u_{N-1}) }{W(u_0,u_1,\ldots,u_{N-1})}
\qquad n=0,\ldots, N-1
\ee
\[
\mathrm{Ker}L^+ = \mathrm{span}\{v_0, \ldots,v_{N-1}\}
\]
where $W_n$ denotes the Wronskian built as a determinant of the
$(N-1)\times (N-1)$ matrix with the $u_n-$related column omitted
$W_n=W(u_0,u_1,\ldots,u_{n-1},u_{n+1},\ldots,u_{N-1})$. The
potential $V_N$ of the Hamiltonian $h_N$ can be expressed as
\cite{CK,Krein}
\begin{equation}\label{V1}
 \nonumber
V_N=V_0-2 \left[\ln W(u_0,u_1,\ldots,u_{N-1})\right]''\,.
\end{equation}

In general, it is not excluded that the transformation operator
$L$ may move a solution of $h_0$ out of $\cH_0=\cL^2(a,b)$ thus
transforming a physical solution of $h_0$ into an unphysical
solution of $h_N$. Moreover the inverse scenario is also possible,
i.e. $L$ may transform an unphysical solution of $h_0$ into a
physical solution of $h_N$. In such  cases the point spectrum of
$h_N$ will differ from that of $h_0$. Subsequently, we will
concentrate on mild transformations $L$, which leave most of the
original spectrum invariant with exception of a finite number of
spectral points --- a characteristic feature of differential
intertwining operators $L$ leaving the boundary behavior of the
solutions of the \Sh equation unchanged. Recently in
\cite{A-S-ZNSP} a conjecture has been proven, which was originally
formulated in \cite{PLA1999} and which states that any $N$th-order
mild differential transformation $L$ can be constructed as a
superposition from only first- and second-order mild
transformations. In this case it is possible to show \cite{BS}
that for problems formulated over the whole $\RR$ (for infinite
values of $a$ and $b$) $V_N$ behaves asymptotically like $V_0$.
Therefore the operator $h_N$ is also essentially self-adjoint and
``lives'' in the same Hilbert space $\cH$ as $h_0$. Moreover,
since the point spectrum of the self-adjoint Sturm-Liouville
problems that we consider is non-degenerate there is no way to
create a new discrete level at the position of an already existing
discrete level and by this means to increase the geometric
multiplicity of that level\footnote{This is in contrast to SUSY
intertwined non-self-adjoint operators $h_0$, $h_2$ for which a
second-order intertwiner $L$ can map two distinct discrete levels
of $h_0$ into a second-order Jordan block of $h_2$, i.e. an
eigenvalue of geometric multiplicity one and algebraic
multiplicity two (for the details see Ref
\cite{BFS-JPA-Lett-2005}).}.

According to \cite{Krein} the necessary condition for an $N$th-order
transformation to produce an essentially self-adjoint operator $h_N$
is%
\footnote{The basic idea can be understood as signature preservation
of Hilbert space metrics, i.e. $\forall E$ belonging to the point
spectrum of $h_0$ the eigenfunctions $\psi_E$ with
$||\psi_E||^2=(\psi_E,\psi_E)> 0$ should map into  corresponding
eigenfunctions $\phi_E$ of $h_N$ with
$||\phi_E||^2=||L\psi_E||^2=(\psi_E,L^+L\psi_E
)=(\psi_E,P_N(h_0)\psi_E)\ge 0$ what via \rf{polyn-1} implies
\rf{usl}. Here, the equality takes place for those $\psi_E$ for
which the point $E$ does not belong to the spectrum of $h_N$, i.e.
if $P_N(E)=0$ and $\phi_E=L\psi_E\equiv0$. A more detailed analysis
of corresponding sufficient conditions for spectral problems on the
whole real line in case of scattering potentials and of confining
potentials is given in \cite{PLA1999}.}
\be\label{usl}
(E-\a_0)\ldots(E-\a_{N-1})\geqslant0
\qquad \forall E\in \mathrm{spec}(h_0)\,.
\ee
This criterion ensures the mildness of the transformation $L$
leading only to changes in maximally $N$ spectral points (of the
point spectrum). Specifically, the spectrum of $h_N$ may contain $p$
points more and $q$ points less than $\mathrm{spec}(h_0)$, where
necessarily $p+q\leqslant N$.

In the present paper we will consider the following possibilities.
\begin{itemize}
\item
The spectrum of $h_0$ is a subset of the spectrum of $h_N$. A new
energy level may be created in the spectrum of $h_N$ if and only if
the corresponding transformation function $u(x)$ is such that
$1/u(x)\in \cD(h_N)$, i.e., in particular, that $1/u(x)$ is
$\cL^2-$integrable and satisfies the Dirichlet BCs.
\item
The spectrum of $h_N$ is a subset of the spectrum of $h_0$. An
energy level may be removed from the spectrum of $h_0$ if and only
if the corresponding transformation function $u(x)$ coincides with
the $h_0-$eigenfunction of this level, i.e. when $u(x)$ satisfies the
Dirichlet BCs.
\item
The spectrum of $h_0$ coincides with the spectrum of $h_N$. In
this case non of the transformation functions $u_l(x)$ nor
$1/u_l(x)$ should be physical, i.e. satisfy Dirichlet BCs on both
ends of the interval $(a,b)$.  This property should be fulfilled
for all transformation functions $u_l(x)$ from which the
transformation operator is built.
\end{itemize}
In all cases we assume that the transformation functions
$\{u_l(x)\}_{l=1}^M$ are linearly independent one from the other and
their Wronskian $W(u_1,\ldots,u_M)$ does not vanish  $\forall
x\in(a,b)$.

In the remainder of this section, we briefly recall some basic
properties of quantum mechanical propagators. Once, the
super-Hamiltonian \eqref{superhamiltonian} is diagonal it suffices
to restrict to one-component (scalar) propagators.

It is well known that the propagator $K(x,y,t)$ of a non-stationary
\Sh equation contains the complete information about the space-time
behavior of the wave function $\Psi(x,t)$ evolving from an initial
configuration $\Psi(x,0)=\Psi_0(x)$
\[
\Psi(x,t)=\int_a^b K(x,y,t)\Psi(y,0)dy
\]
 solving in this way the probability related Cauchy problem of QM.
The propagator (integration kernel) satisfies a differential
equation with Dirac delta function as initial condition
\be\label{K00} [\rmi\p_t-h(x)]K(x,y;t)=0\qquad
K(x,y;0)=\delta(x-y)\,.
\ee
For non-dissipative systems, like in
our case, the propagator $K(x,y;t)$ can be interpreted as
coordinate representation of the unitary evolution operator
$U(t)$: \ \ $K(x,y;t)=\langle x|U(t)|y\rangle $, where unitarity
implies the symmetry $K^*(x,y;-t)=K(y,x;t)$.

Subsequently, we will mainly work with a spectral decomposition of
propagators in terms of complete basis sets of eigenfunctions
\be\label{decomp}
K(x,y;t)= \sum_{n=0}^{N_p}\psi_n(x,t)\psi_n^*(y)+
\int_{-\infty}^\infty dk\,\psi_k(x,t)\psi_k^*(y)
\ee
where summation over the point spectrum and integration over the
continuous (essential) spectrum are understood. By
SUSY-transformations we will only induce changes in the point
spectrum of $h_0$ so that, for simplicity,  we will work with
decompositions over discrete sets of basis functions (corresponding
to point spectra) keeping in mind that extensions to the continuous
spectrum are straight forward.

It is clear that the defining equation for the propagator of the
non-stationary \Sh equation with super-Hamiltonian
\eqref{superhamiltonian} can be trivially  decomposed as
\bea\nonumber
[iI\p_t-H]\bK(x,y;t)=0\qquad
\bK=\left[ \begin {array}{cc}
K_0(x,y;t)&0\\\noalign{\medskip}
0&K_N(x,y;t)\end {array}
 \right]\\
\bK(x,y;0)=I\delta(x-y)\,.
\nonumber
\eea

\section{Propagators related by first-order intertwiners\label{SUSY-1}}

In this section we study the structure of propagators interrelated
by first-order SUSY transformations. Such SUSY transformations are
generated from a single function $u(x)$ and they act as basic
building blocks in chain representations of higher-order
intertwiners. Although any $N$th-order intertwiner may be
represented as a chain of first order intertwiners \cite{BS}, one
has to distinguish between chains which are completely reducible
within a given Hilbert space $\cH$ \cite{MPLA96} and chains which
are partially or completely irreducible in $\cH$. Complete
reducibility means that apart from $h_0$ and $h_N$ also all
intermediate first-order SUSY related Hamiltonians $h_k$, $k\neq
0,N$ are self-adjoint or essentially self-adjoint in the same
Hilbert space $\cH$. In case of irreducible chains\footnote{For a
careful analysis of different kinds of irreducible transformations
we refer to \cite{A-S-ZNSP}.} several or all intermediate
Hamiltonians are non-self-adjoint in $\cH$. Below both chain types
will play a role. We note that chain representations lead to
extremely simplified transformation rules for higher-order
intertwined propagators and allow for very efficient calculation
techniques (see subsection \ref{4.2} below).

According to \eqref{operator} a first-order intertwiner has the
form
\[
L_x=-u'(x)/u(x)+\partial_x\qquad h_0u=\a u\,.
\]
Due to $E-\alpha\ge 0$
(see \eqref{usl})
and depending on the concrete form of the
function $u(x)$ the intertwiner $L$ may result in the following
three types of relations between the spectra of the
Hamiltonians $h_0$ and $h_1$ (see, e.g., \cite{Sukumar1}):\\
{\bf (i)} for $\a=E_0$ and $u=\psi_0$ the ground state level $E_0$
of $h_0$ is
removed from the spectrum of $h_1$,\\
{\bf (ii)} $h_1$ has a new and deeper ground state level
$E_{-1}=\a<E_0$
than $h_0$,\\
{\bf (iii)} the spectra of $h_1$ and $h_0$ completely coincide
($\a<E_0$).\\
In order to create a potential $V_1(x)$ which is nonsingular on the
whole interval $(a,b)\ni x$ the function $u(x)$ should be nodeless
inside this interval. This property is evidently fulfilled for type
(i) relations since the function $u(x)$ coincides in this case with
the ground state eigenfunction $u(x)=\psi_0(x)$. In the cases (ii)
and (iii) the nodelessness should be ensured by an appropriate
choice of $u(x)$, a choice which is always possible because of the
`oscillation' theorem (see e.g. \cite{Berezin}). In case (iii) it
implies $\a<E_0$.

Introducing the Green function
\be\label{G0}
G_0(x,y;E)=\sum_{m=0}^{\infty}\frac{\psi_m(z)\psi_m(y)}{E_m-E}
\ee
of the stationary $h_0-$\Sh equation at fixed energy $E$%
\footnote{We assume that $h_0$ has a purely discrete spectrum. As it
was already stated in the Preliminaries, a generalization to the
continuous spectrum is straight forward. } (see, e.g. \cite{Morse})
and its ``regularized'' version
\[\fl
\widetilde{G}_0(z,y,E_0)=\sum_{m=1}^{\infty}\frac{\psi_m(x)\psi_m(
y)}{E_m-E_0}=\lim_{E\rightarrow
E_0}\left[G_0(z,y,E)-\frac{\psi_0(x)\psi_0(y)}{E_0-E}\right]
\]
the corresponding structural relations for the propagators can be
summarized in the following
\begin{Th}
The propagators $K_1(x,y;t)$ and $K_0(x,y;t)$ of non-stationary
\Sh equations with SUSY intertwined Hamiltonians $h_1$ and $h_0$
are interrelated with each other and with the Green functions
$G_0(x,y;E)$ and $\widetilde{G}_0(z,y,E_0)$
in the following way:\\
Type {\bf (i)} relation
\be\label{trprv1}
K_1(x,y,t)=L_xL_y\int_{a}^{b}K_0(x,z,t)\widetilde{G}_0(z,y,E_0)dz\,.
\ee
Type {\bf (ii)} relation
\be\label{trpr1}\fl
K_1(x,y,t)=L_xL_y\int_{a}^{b}K_0(x,z,t)G_0(z,y,\a)dz+
\phi_{-1}(x)\phi_{-1}(y)\rme^{-\rmi\a t}.
\ee
Type {\bf (iii)} relation
\be\label{trpris2}
K_1(x,y,t)=L_xL_y\int_{a}^{b}K_0(x,z,t)G_0(z,y,\a)dz\,.
\ee
\end{Th}

\begin{proof}
We start from the type {\bf (ii)} relation and represent the
propagator $K_1(x,y;t)$ in terms of the basis functions
${\phi_m(x,t)}$ of the Hamiltonian $h_1$ (cf. \eqref{decomp}). We
note that the explicit time-independence of $h_1$ implies a
factorization
$\phi_m(x,t)=\phi_m(x)\exp({-\rmi E_mt})$ with $\phi_m(x)$
purely real-valued. Expressing $\phi_m$ in terms of the
corresponding wave functions of the Hamiltonian $h_0$,
$\phi_m=N_mL\psi_m$, with  $N_m=(E-\a)^{-1/2}$ a normalization
constant (see, e.g. \cite{Sukumar1}), we arrive at
\begin{eqnarray*}
K_1(x,y,t)&=&\sum_{m=-1}^{\infty}
\phi_m(x)\phi_m(y)\rme^{-\rmi E_mt}\nn\\
&=& L_xL_y\sum_{m=0}^{\infty}
\frac{\psi_m(x)\psi_m(y)}{E_m-\a}\rme^{-\rmi E_mt}+
\phi_{-1}(x)\phi_{-1}(y)\rme^{-\rmi\a t}\,.
\end{eqnarray*}
Modulo a normalization factor $N$, the wave function $\phi_{-1}$
of the new ground state is
proportional to the inverse power of the
 transformation
function $u(x)$, $\phi_{-1}=N/u(x)$. It remains to express the
time-dependent phase factor in terms of the propagator. This can be
easily done using the evident property of the bound state solutions
of the \Sh equation
\be\label{vc0-1}
\int_{a}^{b}K_0(x,z,t)\psi_m(z)dz=\psi_m(x)\rme^{-\rmi E_mt}
\ee
so that the previous equation  reads
\be\label{vc1} \fl
K_1(x,y,t)=
L_xL_y\int_{-\infty}^{\infty}K_0(x,z,t)
\sum_{m=0}^{\infty}\frac{\psi_m(z)\psi_m(y)}{E_m-\a}dz
+ \phi_{-1}(x)\phi_{-1}(y)\rme^{-\rmi\a t}.
\ee
The sum in this relation can be identified as the Green function
\eqref{G0}.
Due to $E_m-\alpha>0, \ \forall E_m\in\mathrm{spec}(h_0)$ this Green
function is regular $\forall E_m$ and the proof for type {\bf (ii)}
transformations is complete.

The proof for type {\bf (i)} and {\bf (iii)} transformations follows
the same scheme. The formally regularized Green function
$\widetilde{G}_0(z,y,E_0)$ in {\bf (i)} results from the fact that
the ground state with energy $E_0$ is not present in the spectrum of
$h_1$ so that a sum $\sum_{m>0}$ appears and the ground state
contribution has to be subtracted from $G_0(x,y,\a=E_0)$. In case of
a type  {\bf (iii)} transformation a sum $\sum_{m=0}^\infty $ over
the complete set of eigenfunctions appears in \eqref{vc1} and no new
state occurs.
\end{proof}

We conclude this section by reshaping relation \eqref{trprv1} for
the propagator of a system with removed original ground state, i.e.
of a type {\bf (i)} transformed system. The corresponding result can
be formulated as
\begin{Th}\label{th2}
For transformations with $u(x)=\psi_0(x)$ the
propagator $K_1(x,y;t)$ of the resulting system can be represented
as
\begin{equation}\label{trans1l}\fl
 K_1(x,y;t)=-\frac{1}{u(y)}L_x\int_{a}^y
   K_0(x,z;t)u(z)dz=\frac{1}{u(y)}L_x\int_{y}^b
   K_0(x,z;t)u(z)dz\,.
\end{equation}
\end{Th}
First of all we recall that $\psi_0(x)$ being the ground state
function of $h_0$ satisfies the zero boundary conditions. To
facilitate the proof of Theorem \ref{th2} we need the following two
lemmas.
\begin{Lm}\label{lemma1}
\[\fl\quad
L_y\lim_{E\rightarrow E_0}
\left(G_0(z,y,E)-\frac{\psi_0(z)\psi_0(y)}{E_0-E}\right)
=\lim_{E\rightarrow E_0}L_yG_0(z,y,E)\,.
\]
\end{Lm}
\begin{proof}
This result follows from the explicit
representation of $G_0(z,y,E)$ in terms of basis functions.
On the
one hand, it holds
\[
L_y\lim_{E\rightarrow
E_0}\left(G_0(z,y,E)-\frac{\psi_0(z)\psi_0(y)}{E_0-E}\right)
=\sum_{n=1}^{\infty}\frac{\psi_n(z)L_y\psi_n(y)}{E_n-E_0}
\]
whereas on the other hand the kernel property (annihilation) of
the ground state $L\psi_0=0$ gives
\[\fl
\lim_{E\rightarrow E_0}\left(L_yG_0(z,y,E)\right)=
\lim_{E\rightarrow
E_0}\left(\sum_{n=1}^{\infty}\frac{\psi_n(z)L_y\psi_n(y)}{E_n-
E}\right)
=\sum_{n=1}^{\infty}\frac{\psi_n(z)L_y\psi_n(y)}{E_n-E_0}\,.
\]
\end{proof}

\begin{Lm}\label{lemma2}
Let $f_l(x,E)$ and $f_r(x,E)$ satisfy the \Sh equation
\be\label{eqf}\fl h_0f(x,E):=-f''(x,E)+V_0(x)f(x,E)=Ef(x,E)\qquad
x\in(a,b)
\ee
and boundary conditions
\be \label{bcf}
f_l(a,E)=0\qquad f_r(b,E)=0\,.
\ee
Let also $E=E_0$ be the ground state level of
$h_0$ with $\psi_0(x)$ as the ground state function
(we assume that $h_0$ has at least one discrete level) then
\bea\label{lim1}
 \lim_{E\rightarrow E_0}\frac{f_l(x,E)L_yf_r(y,E)}{W(f_r,f_l)}
&=&-\frac{\psi_0(x)\int_y^b \psi_0^2(z)dz}{\psi_0(y)\int_a^b
\psi_0^2(z)dz}
\\[.5em]
 \lim_{E\rightarrow E_0}\frac{f_r(x,E)L_yf_l(y,E)}{W(f_r,f_l)}
&=&\frac{\psi_0(x)\int_a^y \psi_0^2(z)dz}{\psi_0(y)\int_a^b
\psi_0^2(z)dz}
 \label{lim2}
\eea
where $L_y=-u'(y)/u(y)+\p_y$ with $u(y)\equiv\psi_0(y)$.
\end{Lm}

\begin{proof}
First we note that according to \eqref{L1} $L_yf_r(y,E)=W(u,f_r)/u$.
Next, since both $u=\psi_0$ and $f_r$ satisfy the same \Sh equation
(\ref{eqf}) it holds $W'(u,f_r)=(E_0-E)uf_r$ and hence
\be\label{l2-1}
W(u,f_r)=(E-E_0)\int_y^bu(z)f_r(z,E)dz\
\ee
where we have used the property $W(u,f_r)_{y=b}=0$ which follows
from the BCs for $u$ and $f_r$. Via \rf{l2-1} we find
\be
L_yf_r(y,E)=\frac{E-E_0}{u(y)}\int_y^bu(z)f_r(z,E)dz
\ee
and hence
\be\label{LFW}
\frac{L_yf_r(y,E)}{W(f_r,f_l)}=-\frac{E-E_0}{f_l(b,E)}\
\frac{\int_y^bu(z)f_r(z,E)dz}{f'_r(b,E)u(y)}
\ee
where it has
been used that the Wronskian
$W(f_r,f_l)=f_r(x,E)f'_l(x,E)-f_l(x,E)f'_r(x,E)$ is $x$-independent
and can be calculated at $x=b$ where $f_r(b,E)=0$. Since the
spectrum of $h_0$ is non-degenerate, the ground state function is
unique up to an arbitrary constant factor and, hence,
$u(x)=\psi_0(x)$, $f_r(x,E_0)$ and $f_l(x,E_0)$ have to be
proportional to each other
\be\label{fC} f_{r,l}(x,E_0)=C_{r,l}u(x) \ee
and for $E\to E_0$  only the first fraction in \eqref{LFW} remains
undetermined. The l'Hospital rule gives for this limit
\be\label{lEE}
\lim_{E\to E_0}\frac{E-E_0}{f_l(b,E)}=
\frac{1}{\dot f_l(b,E_0)}
\ee
where the dot denotes the derivative with respect
to $E$. Making use of \rf{lEE} and
\be\label{ff} \dot{f_l}(b,E_0)f_l'(b,E_0)=\int_a^b f_l^2(z,E_0)dz
\ee
(which we  prove below) relation \eqref{LFW} yields
\be
\fl
 \lim_{E\rightarrow E_0}\frac{f_l(x,E)L_yf_r(y,E)}{W(f_r,f_l)}
=-f_l(x,E_0)\frac{f'_l(b,E_0)}{f'_r(b,E_0)}\
\frac{\int_y^bu(z)f_r(z,E_0)dz}{u(y)\int_a^bf^2_l(z,E_0)dz}
\ee
and via \eqref{fC} it leads to the result \eqref{lim1}. The proof of
\eqref{lim2} follows the same lines with evident changes.

Finally, it remains to derive equation \eqref{ff}. This is easily
accomplished by multiplying \Sh equation \eqref{eqf} for
$f=f_l(x,E)$ by $\dot f_l(x,E)$, its derivative with respect to $E$
by $f=f_l(x,E)$, and integrating their difference over the interval
$(a,b)$. The intermediate result
\bea
\int_a^bf_l^2(x,E)dx& & \nonumber
\\ \fl
=\dot f'_l(a,E)f_l(a,E)-f'_l(a,E)\dot f_l(a,E) -\dot
f'_l(b,E)f_l(b,E)+f'_l(b,E)\dot f_l(b,E) & &
\eea
reduces to \eqref{ff} via BC \eqref{bcf} and its derivative with
respect to $E$ (what cancels the first two terms) and the limit
$E=E_0$, its implication \eqref{fC} and the BC for $u(x)$.
\end{proof}

\noindent {\it Proof of theorem \ref{th2}.} For the Green function
$G(x,y,E_0)$ in \eqref{trprv1} we use the standard representation in
terms of two linearly independent solutions $f_{l,r}$ of the
$h_0-$\Sh equation introduced in Lemma \ref{lemma2}
 (see, e.g. \cite{Levitan}):
\be \fl
G(x,y,E)=\left[f_l(x,E)f_r(y,E)\T(y-x)+
f_l(y,E)f_r(x,E)\T(x-y)\right]/W(f_r,f_l)
 \ee
where $\T$ denotes,
as usual, the Heavyside step function. Then relation \eqref{trprv1}
takes the form
\begin{eqnarray*}\fl
&&K_1(x,y,t)=\nn\\
\fl &&L_xL_y\!\!\int_{a}^{b}\!\!K_0(x,z,t)
\lim_{E\rightarrow E_0}\!
\left[\frac{f_l(z)f_r(y)}{W(f_r,f_l)}
\Theta(y-z)+\frac{f_l(y)f_r(z)}{W(f_r,f_l)}\Theta(z-y)
-\frac{\psi_0(z)\psi_0(y)}{E_0-E}\right]dz
\end{eqnarray*}
where the step functions can be resolved to give
\bea \nonumber \fl
K_1(x,y,t)&=& L_xL_y\int_{a}^{y}K_0(x,z,t)\lim_{E\rightarrow
E_0}\left[\frac{f_l(z)f_r(y)}{W(f_r,f_l)}
-\frac{\psi_0(z)\psi_0(y)}{E_0-E}\right]dz
\\ \nonumber
\fl &+&L_xL_y\int_{y}^{b}K_0(x,z,t)\lim_{E\rightarrow
E_0}\left[\frac{f_l(y)f_r(z)}{W(f_r,f_l)}
-\frac{\psi_0(z)\psi_0(y)}{E_0-E}\right]dz\,.
\eea
The second argument of the functions $f_{l,r}$ has been omitted for
notational simplicity. Explicitly acting with the differential
operator $L_y$ on the integrals with variable $y-$boundary yields
\bea\fl
\nonumber K_1(x,y,t)&=&
L_x\int_{a}^{y}K_0(x,z,t)L_y\lim_{E\rightarrow
E_0}\left[\frac{f_l(z)f_r(y)}{W(f_r,f_l)}
-\frac{\psi_0(z)\psi_0(y)}{E_0-E}\right]dz
\\ \fl
&+&L_x\int_{y}^{b}K_0(x,z,t)L_y\lim_{E\rightarrow
E_0}\left[\frac{f_l(y)f_r(z)}{W(f_r,f_l)}
-\frac{\psi_0(z)\psi_0(y)}{E_0-E}\right]dz
\eea
whereas via Lemma \ref{lemma1}  the intertwiner $L_y$ and the limit
$\lim_{E\rightarrow E_0}$ can be interchanged to give
\bea\label{prop1l} \fl &&K_1(x,y,t)=\\
\fl && \nonumber L_x\left\{\int_{a}^{y}K_0(x,z,t)\lim_{E\rightarrow
E_0}\frac{f_l(z)L_yf_r(y)}{W(f_r,f_l)}dz+
\int_{y}^{b}K_0(x,z,t)\lim_{E\rightarrow
E_0}\frac{f_r(z)L_yf_l(y)}{W(f_r,f_l)}dz\right\}\,.
\eea
Application of Lemma \ref{lemma2} leads to
\bea\label{prop2l}
\nonumber\fl
 K_1(x,y,t)=
\frac{L_x}{\psi_0(y)\int_a^b \psi_0^2(q)dq} \left\{- \int_y^b
\psi_0^2(q)dq \int_{a}^{y}K_0(x,z,t)\psi_0(z) dz \right.
\\\hspace{-2em}
+\left. \int_a^y \psi_0^2(q)dq
\int_{y}^{b}K_0(x,z,t)\psi_0(z)dz\right\}
\eea
which we further reshape by expressing the integral with respect to
$q$ over the interval $(y,b)$ by the difference of two integrals
over the intervals $(a,b)$ and $(a,y)$. Substitution of $\psi_0=u$
in the first term results in
\bea\label{prop3l}  \nonumber
 \fl K_1(x,y,t)=-
\frac{1}{u(y)}L_x \int_{a}^{y}K_0(x,z,t)u(z) dz
\\\hspace{-2em}
+ \frac{1}{\psi_0(y)\int_a^b \psi_0^2(q)dq} \int_a^y \psi_0^2(q)dq\
L_x \int_{a}^{b}K_0(x,z,t)\psi_0(z)dz\,.
\eea
The very last integral is nothing but the ground state stationary
wave function $\psi_0(x,t)=u(x)\exp(-\rmi E_0t)$. Therefore, since
$L_xu(x)=0$, we obtain the first equality in \eqref{trans1l}. The
second equality results from applying a similar transformation to
the second term in \eqref{prop2l}. \hfill $\square$

The following remarks are in order. First we have to note that the
integral representation \eqref{trans1l} is only valid in case of
first-order SUSY transformations which remove the ground state
level. If one wants to create a level in a problem on the whole real
line one has to use a transformation function $u(x)$ which diverges
for $x\to\pm\infty$ ensuring in this way the normalizability and
Dirichlet BCs of the new ground state wave function
$\phi_{-1}(x)\propto 1/u(x)$. An attempt to calculate the propagator
$K_1$ via \rf{trans1l} would usually lead to a divergent integral.
The correct approach is to use \eqref{trpr1} in this case. Jauslin
\cite{jauslin} using a different procedure obtained the same result
\eqref{trans1l} both for removing and creating a level, but he
completely ignored questions of convergence or divergence of the
corresponding integrals. In concrete calculations he avoided
divergent integrals by considering the heat equation only.

\section{Higher order transformations\label{SUSY-N}}

\subsection{Addition of new levels\label{4.1}}

Let us consider an $N$th-order ($N=2,3,\ldots$) polynomial
supersymmetry corresponding to the appearance of $N$ additional
levels in the spectrum of $h_N$ compared to the spectrum of $h_0$.
In this case new levels may appear both below the ground state
energy of $h_0$ (reducible supersymmetry) and between any two
neighbor levels of $h_0$ (irreducible supersymmetry, see e.g.
\cite{PLA1999}). The propagator for the transformed equation can be
found in the following way. We develop $K_N(x,y;t)$ over the
complete orthonormal set $\{\phi_m(x,t)\}$ of eigenfunctions of
$h_N$ and express all $\phi_m$ with eigenvalues already contained in
the spectrum of $h_0$ in terms of $\psi_m$, i.e.
$\phi_m=N_mL\psi_m$. The normalization constants $N_m$ for
transformations fulfilling condition \rf{usl} have the form
\cite{BS}:
\[N_m=[(E-\a_0)(E-\a_1)\ldots (E-\a_{N-1})]^{-1/2}\,.\]
All other eigenfunctions of $h_N$ which correspond to new levels and
which are not contained in $\spec (h_0)$ we keep untouched. This
yields
\[\fl\label{prop-4-1}
K_N(x,y,t)=
L_xL_y\sum_{m=0}^{\infty}
\frac{\psi_m(x)\psi_m(y)}{(E_m-\a_0)\ldots(E_m-\a_{N-1})}
\rme^{-\rmi E_mt}+
\sum_{n=0}^{N-1}\phi_n(x)\phi_n(y)\rme^{-\rmi\a_nt}\,.
\]
Here we interchanged the  derivative operators present in $L_{x,y}$
and the summation. This interchange is justified because the
propagators are understood not as a usual functions but as
generalized functions \cite{GelShil}
(which, in particular, may be regular, i.e. defined
with the help of locally integrable functions). It remains to
express $\psi_m(x)\exp({-\rmi E_mt})$ with the help of \rf{vc0-1} in
terms of $K_0(x,z,t)$, to make use of the identity
\[
\prod_{n=0}^{N-1}\frac{1}{E-\a_n}=\sum_{n=0}^{N-1}
\left(\prod_{j=0,j\neq n}^{N-1}
\frac{1}{\a_j-\a_n}\right)\frac{1}{E-\a_n}
\]
and to represent the sum over $m$ in terms of the Green function
$G_0(z,y,\a_n)$.
As a result, one arrives at
\bea\nonumber
&& \fl K_N(x,y,t)= L_xL_y\sum_{n=0}^{N-1} \left( \prod_{j=0,j\neq
n}^{N-1}\frac{1}{\a_j-\a_n}\right)
\int_{-\infty}^{\infty}K_0(x,z,t)G_0(z,y,\a_n)dz
\\ \label{trprn}
&+&
\sum_{n=0}^{N-
1}\phi_n(x)\phi_n(y)\rme^{-\rmi\a_nt}\,.
\eea

\subsection{Removal of levels \label{4.2}}

In this section we need information on the intermediate
transformation steps of $N$th-order SUSY-transformations which
goes beyond that presented in section \ref{prelim}. Therefore we
start with a more detailed description of the transformation
operators and solutions of the \Sh equation at each transformation
step.

Let us consider a chain of  $N$ first-order transformations
\[
h_N \stackrel{L_{N,N-1}}{\longleftarrow} h_{N-1} \stackrel{L_{N-1,N-
2}}{\longleftarrow}\ldots\stackrel{L_{2,1}}{\longleftarrow}
h_1\stackrel{L_{1,0}}{\longleftarrow} h_0
\]
built from operators $L_{k+1,k}$ which intertwine neighbor
Hamiltonians $h_k$ and $h_{k+1}$ as
$L_{k+1,k}h_k=h_{k+1}L_{k+1,k}$. We assume all Hamiltonians $h_k$,
$k=0,\ldots, N$ self-adjoint or essentially self-adjoint in the
same Hilbert space $\cH$ so that the SUSY-transformation chain
itself is completely reducible. Furthermore, we assume that at
each transformation step the ground state of the corresponding
Hamiltonian is removed. This means that after $N$ linear
SUSY-transformations the first $N$ states of the $h_0-$system are
removed and the $N+1$st state of $h_0$ maps into the ground state
of $h_N$. In our analysis these first $N+1$ states of $h_0$ will
play a crucial role and we denote them by $u_{0,n}$, $n=0,\ldots,
N$. Furthermore, we use a numbering  for the solutions $u_{k,n}$
of the \Sh equations of the SUSY-chain Hamiltonians $h_k$,
$k=0,\ldots,N$ which is ``synchronized" with the level numbering
of $h_0$, meaning that a function $u_{k,n}$ is related to the
spectral parameter $E_n$. We have to distinguish between physical
solutions, which correspond to the existing bound states of $h_k$
and which have indices $n=k,\ldots, N$, and unphysical auxiliary
solutions $u_{k,n}$ with $n=0,\ldots, k-1$ which we construct
below. The ground state eigenfunction of a Hamiltonian $h_k$ is
given by $u_{k,k}$ and for $k<N$ it is annihilated by the
SUSY-intertwiner $L_{k+1,k}$
\[L_{k+1,k}u_{k,k}=0\,.\]
The bound state functions  $u_{k+1,n}$, $n=k+1,\ldots,N$ of
$h_{k+1}$ may be obtained by acting with the SUSY-intertwiner
\be\label{w1} 
L_{k+1,k}=-u_{k,k,x}/u_{k,k}+\partial_x\qquad
L_{k+1,k}f=\frac{W(u_{k,k},f)}{u_{k,k}}
\ee
on the corresponding eigenfunctions of $h_k$
\[
L_{k+1,k}u_{k,n}=u_{k+1,n}\qquad n=k+1,\ldots, N\,.
\]

Next, we note that the chain of $k$, ($k=2,\ldots,N$) first-order
transformations is equivalent to a single $k$th-order transformation
\eqref{operator} generated by the transformation functions
$u_{0,0},u_{0,1},\ldots,u_{0,k-1}$. Furthermore, the transformation
operators obey the composition rules
\be\label{Lcomposition}
L_{k+1,\,k}L_{k,\,l}=L_{k+1,\,l}\qquad
l=0,\ldots,k-1 \qquad k=1,\ldots,N-1
\ee
so that, e.g., the second-order transformation operator $L_{k+2,k}$
intertwines the Hamiltonians $h_k$ and $h_{k+2}$
\[
L_{k+2,k}h_k=h_{k+2}L_{k+2,k}\qquad k=0,\ldots,N-2\,.
\]
The $N$th-order transformation operator $L_{N,0}$ is then
inductively defined as $L_{N,0}=L_{N,N-1}L_{N-1,0}$\,. Obviously, it
annihilates the $N$ lowest states of the original Hamiltonian $h_0$,
i.e. $u_{0,0},\ldots,u_{0,N-1}\in\mbox{Ker}L_{N,0}$.

As further ingredient for the derivation of the propagator-mapping
we need the set of unphysical auxiliary functions $u_{N,n}$,
$n=0,\ldots, N-1$. We construct them as $u_{N,n}=L_{N,0}\wt
u_{0,n}$\,, where the functions $\wt u_{0,n}$ are the unphysical
solutions of the $h_0-$\Sh equation at energies $E_n$ which are
linearly independent from the eigenfunctions $u_{0,n}$. Normalizing
$\wt u_{0,n}$  by the condition $W(u_{0,n},\tilde u_{0,n})=1$ and
integrating this Wronskian gives
\[
\widetilde{u}_{0,n}(x)=
u_{0,n}(x)\int_{x_0}^x\frac{dy}{u_{0,n}^2(y)}
\]and finally
\be\label{LN0u}
u_{N,n}(x)= L_{N,0}u_{0,n}(x)\int_{x_0}^x\frac{dy}{u_{0,n}^2(y)}
\qquad n=0,\ldots,N-1\,.
\ee
In \ref{ap1} we show that
\bea\label{norm21}\fl
u_{N,n}=C_{N,n}
\frac{W_{n}(u_0,\ldots,u_{N-1})}{W(u_0,\ldots,u_{N-1})}
 \\ \fl \nonumber
C_{N,n}=(E_{N-1}-E_n)(E_{N-2}-E_n)\ldots(E_{n+1}-E_n) \qquad
n=0,\ldots,N-2
\\ \fl
u_{N,N-1}=
\frac{W_{N-1}(u_0,\ldots,u_{N-1})}{W(u_0,\ldots,u_{N-1})}=
\frac{W(u_0,\ldots,u_{N-2})}{W(u_0,\ldots,u_{N-1})}\,.
\label{norm21a}
\eea
The ground state function of $h_N$ is obtained by  acting with
$L_{N,0}$ on the $N$th excited state of $h_0$:
\be\label{uNN}
u_{N,N}=L_{N,0}u_{0,N}=\frac{W(u_{0,0},\ldots,u_{0,N})}%
{W(u_{0,0},\ldots,u_{0,N-1})}\,.
\ee
Below we will often use the abbreviations $W(x)$, $W_n(x)$ for the
Wronskians (when there is no ambiguity in the definition of the
transformation functions $u_{0,m}$ in their arguments) indicating
explicitly only the dependence on the spatial coordinate.

Within the above framework, the $N$ first discrete levels $E_0,
E_1,\ldots,E_{N-1}$ have been removed from the spectrum of $h_0$
by choosing the ground state functions $u_{k,k}$ of the
Hamiltonians $h_k$ as intermediate transformation functions. For
such a construction the transformed propagator may be calculated
according to
\begin{Th}\label{th3}
Let the $N$ first eigenfunctions $u_{0,n}\equiv u_n=\psi_n$,
$n=0,\ldots,N-1$ of $h_0$ be the SUSY transformation functions. Then
the propagators $K_N(x,y;t)$ and $K_0(x,y;t)$ of the \Sh equations
with Hamiltonians $h_N$ and $h_0$ are interrelated as
\bea\label{transNl}\fl
K_N(x,y;t)&=&(-1)^{N}L_{N,0,x}
\sum_{n=0}^{N-1}(-1)^{n}
\frac{W_{n}(y)}{W(y)}
\int_{a}^y   K_0(x,z;t)u_n(z)dz\\
   \fl
  & =& (-1)^{N-1}L_{N,0,x}
  \sum_{n=0}^{N-1} (-1)^{n}
  \frac{W_{n}(y)}{W(y)}
  \int_{y}^b K_0(x,z;t)u_n(z)dz\,.
  \label{transN2}
\eea
\end{Th}

\begin{proof} The proof of these relations can be given by induction.
We start with (\ref{transN2}). For $K_1(x,y;t)$ the statement is
proven in \eqref{trans1l}. Assuming that (\ref{transN2}) holds for
$K_N(x,y;t)$ we verify its validity for $K_{N+1}(x,y;t)$. The
corresponding Hamiltonians $h_{N+1}$ and $h_N$ are intertwined by
the {\em linear} transformation $L_{N+1,N}$ so that \eqref{trans1l}
is applicable and $K_{N+1}(x,y;t)$ can be represented as
\[
K_{N+1}(x,y;t)=\frac{1}{u_{N,N}(y)}L_{N+1,N,x}\int_y^b
K_{N}(x,z;t)u_{N,N}dz\,.
\]
Replacing $K_N$ by (\ref{transN2}) and making use of relations
\eqref{norm21}, \eqref{norm21a} and the composition rule
\eqref{Lcomposition} gives
\bea\nonumber
&&(-1)^{N-1} K_{N+1}(x,y;t)=\frac{1}{u_{N,N}(y)}L_{N+1,0,x}\\
 &&
\times\sum_{n=0}^{N-1} (-1)^{n}C_{Nn}^{-1}
\int_y^bdz\,\int_{z}^bdq\,u_{N,n}(z)u_{N,N}(z)
 K_0(x,q;t)u_{0,n}(q)\,.
 \label{propInt}
\eea The integration region of the double integral is the upper triangle of the
rectangle $y<z,q<b$ in the $(z,q)-$plane. We replace this double
integral by the difference of two double integrals over the whole
rectangle and the lower triangle, respectively,
\bea\fl \nonumber (-1)^{N-1}K_{N+1}(x,y;t)=
\frac{1}{u_{N,N}(y)}L_{N+1,0,x}\sum_{n=0}^{N-1} (-1)^{n}C_{Nn}^{-1}\\
\nonumber \times \left[ \int_{y}^bdz\,
   K_0(x,z;t)u_{0,n}(z)\int_y^bdq\,u_{N,n}(q)u_{N,N}(q)\right.
\\
\left.-\int_{y}^bdz\,
   K_0(x,z;t)u_{0,n}(z)
\int_z^bdq\, u_{N,n}(q)u_{N,N}(q)\right].\label{44}
\eea
Here, we reshape the two integrals of the type $\int_\xi^b dq\,
u_{N,n}(q)u_{N,N}(q)$ as follows. First we note that $u_{N,n}$ and
$u_{N,N}$ are solutions of the same \Sh equation with Hamiltonian
$h_N$ and therefore
\bea
\int_\xi^b dq\, u_{N,n}(q)u_{N,N}(q)
&=&\frac{W\left[u_{N,N}(\xi),u_{N,n}(\xi)\right]}{E_n-E_N}-W_{b,n}\nn\\
&=& \frac{u_{N,N}L_{N+1,N}u_{N,n}}{E_n-E_N}-W_{b,n}
\label{intuu-2}
\eea
where $W_{b,n}:=W\left[u_{N,N}(b),u_{N,n}(b)\right]/(E_n-E_N)$ and
where the second equality was obtained via \eqref{w1}. Applying
the general relation \eqref{norm21} to
$L_{N+1,N}u_{N,n}=u_{N+1,n}$ leads finally to
\bea\label{aux-int-2}
\int_\xi^b dq \, u_{N,n}(q)u_{N,N}(q)
&=&-C_{N,n}u_{N,N}\frac{W_{n}(u_0,\ldots,u_N)}{W(u_0,\ldots,u_N)}-W_{b,n}\nn\\
&=:&-C_{N,n}u_{N,N}(\xi)\frac{W_{n}(\xi)}{W(\xi)}-W_{b,n}\,.
\eea
With \rf{aux-int-2} as substitution rule the propagator \eqref{44}
takes the form
\bea\label{spA} \fl   \nonumber
(-1)^{N-1}K_{N+1}(x,y;t)=-L_{N+1,0,x}\sum_{n=0}^{N-1}
(-1)^{n}\frac{W_{n}(y)}{W(y)} \int_{y}^b
   K_0(x,z;t)u_{0,n}(z)dz
\\ \fl +\frac{1}{u_{N,N}(y)}L_{N+1,0,x}\int_{y}^bdz
   K_0(x,z;t)\frac{u_{N,N}(z)}{W(z)}\sum_{n=0}^{N-1}
(-1)^{n}u_{0,n}(z)W_n(z) \,.
\eea
(The terms containing $W_{b,n}$ exactly cancelled.) The sum
\be\label{xx1}
S_N:=\sum_{n=0}^{N-1} (-1)^{n}u_{0,n}W_n(u_0,\ldots,u_N)
\ee
in the second term can be calculated explicitly. Comparison with
the evident determinant identity
\bea\nonumber
0=
\left|\begin{array}{cccc}u_{0,0} &  \ldots & u_{0,N-1}&u_{0,N}\\
u_{0,0} &  \ldots & u_{0,N-1}&u_{0,N}\\
\ldots&\ldots&\ldots&\ldots\\
u_{0,0}^{(N-1)} &  \ldots & u_{0,N-1}^{(N-1)}&u_{0,N}^{(N-1)}\\
\end{array}\right|
\\ \fl
=
\left|\begin{array}{cccc}u_{0,0} &  \ldots & u_{0,N-1}&0\\
u_{0,0} &  \ldots & u_{0,N-1}&u_{0,N}\\
\ldots&\ldots&\ldots&\ldots\\
u_{0,0}^{(N-1)} &  \ldots & u_{0,N-1}^{(N-1)}&u_{0,N}^{(N-1)}\\
\end{array}\right|+
\left|\begin{array}{cccc}0 &  \ldots & 0&u_{0,N}\\
u_{0,0} &  \ldots & u_{0,N-1}&u_{0,N}\\
\ldots&\ldots&\ldots&\ldots\\
u_{0,0}^{(N-1)} &  \ldots & u_{0,N-1}^{(N-1)}&u_{0,N}^{(N-1)}\\
\end{array}\right|
\label{xx3}
\eea
shows that \eqref{xx1} coincides with the decomposition of the
first determinant in the second line of \eqref{xx3} over the
elements of its first row.  Hence, it holds
\[
S_N=-(-1)^Nu_{0,N} W(u_{0,0},\ldots,u_{0,N-1})\,.
\]
Representing the ground state eigenfunctions $u_{N,N}$ in
\rf{spA} via \eqref{uNN} in terms of Wronskian fractions we find
that
\[
\frac{u_{N,N}(z)}{W(z)}\sum_{n=0}^{N-1}
(-1)^{n}u_{0,n}(z)W_n(z)=-(-1)^N u_{0,N}(z)
\]
and, hence, that the second term in \eqref{spA} is nothing but the
absent $n=N$ summand of the sum in the first term.
As a result, we
arrive at
\[\fl
K_{N+1}(x,y;t)=(-1)^{N}L_{N+1,0,x}\sum_{n=0}^{N}
(-1)^{n}\frac{W_{n}(u_0,\ldots,u_N)}{W(u_0,\ldots,u_N)} \int_{y}^b
   K_0(x,z;t)u_{0,n}(z)dz
   \]
what completes the proof of \eqref{transN2}. The representation
\eqref{transNl} follows from \eqref{transN2} and  the relations
\[
    \int_{a}^{b}K_0(x,z,t)u_n(z)dz
    =u_n(x)\exp(-\rmi E_nt) \qquad
    L_{N,0,x}u_n(x)=0\,.
\]
\end{proof}
We note that in formulas (\ref{transNl}) and \eqref{transN2} only
one-dimensional integrals are present. In this way, they may turn
out more convenient for concrete calculations than similar
equations derived in \cite{jauslin}.

Furthermore, we note the following. Theorem \ref{th3} is proven
for the case when the $N$ lowest discrete levels are removed from
the spectrum of $h_0$ starting from the ground state level. This
scenario corresponds to reducible supersymmetry. In order to see
which of the conditions on the transformation functions $u_{0,n}$
used for the construction of the propagator representations
\eqref{transNl} and \eqref{transN2} are indeed necessary
conditions one may simply insert $K_N(x,y;t)$ directly into \Sh
equation \eqref{K00}. It turns out that neither the condition of
level deletion starting from the ground state nor a deletion of a
level block without surviving levels inside is used. This means
that equations \eqref{transNl} and \eqref{transN2} hold for any
choice of transformation functions provided their Wronskian does
not vanish inside the interval $(a,b)$, i.e. it holds for
reducible as well as for irreducible SUSY transformation chains. A
necessary but in general not sufficient condition for the
nodelessness of the Wronskian is inequality \eqref{usl} (for
further details see \cite{PLA1999}).

\subsection{Strictly isospectral transformations\label{comp-iso}}

Strictly isospectral transformations can be generated with the help
of unphysical solutions of the \Sh equation as transformation
functions. In this section, we extend theorem \ref{th3} to models
defined over the whole real line $(a,b)=(-\infty,\infty)$ by
assuming transformation functions which vanish at one of the
infinities $x\to-\infty$ or $x\to \infty$ and violate the Dirichlet
BCs at the opposite infinities ($x\to \infty$ or $x\to -\infty$).

In accordance with \eqref{transNl} and \eqref{transN2} we
formulate the corresponding relaxed version of theorem \ref{th3}
as
\begin{Th}\label{th3a}
Let the transformation functions $u_n(x)$ vanish at only one of the
infinities $x\to-\infty$ or $x\to \infty$ of the real axis $\RR$.
Then the propagators $K_N(x,y;t)$ and $K_0(x,y;t)$ of the \Sh
equations with $h_N$ and
$h_0$ as Hamiltonians are related as follows:\\
\[\fl\for u_n(x\to -\infty)\to 0: \]
\begin{equation}\label{transNll}\fl
K_N(x,y;t)=(-1)^N L_x\sum_{n=0}^{N-1}(-1)^{n}
\frac{W_{n}(y)}{W(y)}\int_{-\infty}^y
   K_0(x,z;t)u_n(z)dz
\end{equation}
\[\fl\for u_n(x\to\infty)\to 0: \]
\begin{equation}\label{transNlr}\fl
K_N(x,y;t)= (-1)^{N-1} L_x\sum_{n=0}^{N-1}
(-1)^{n}\frac{W_{n}(y)}{W(y)}\int_{y}^\infty
   K_0(x,z;t)u_n(z)dz
\end{equation}
\[\fl\for u_k(x\to-\infty)\to 0\quad k=0,\ldots,M \ {\rm{and}}\ u_m(x\to\infty)\to 0
\quad m=M+1,\ldots,N-1:\]
\bea\label{transNllr}\fl  \nonumber
K_N(x,y;t)&=&(-1)^NL_x\sum_{k=0}^{M}(-1)^{n}
\frac{W_{k}(y)}{W(y)}\int_{-\infty}^y
   K_0(x,z;t)u_k(z)dz\\
\fl  &&+ (-1)^{N-1}L_x\sum_{m=M+1}^{N-1}
(-1)^{m}\frac{W_{m}(y)}{W(y)}\int_{y}^\infty
   K_0(x,z;t)u_m(z)dz\,.
\eea
\end{Th}
\begin{proof}
We have to verify that the initial condition
$K_0(x,y,0)=\delta(x-y)$ and the \Sh equations
$(\rmi\p_t-h_{0x})K_0(x,y,t)=0$ and $(\rmi\p_t-h_{0y})K_0(x,y,t)=0$
fulfilled by the original propagator $K_0(x,y,t)$ map into
corresponding relations for the final propagator $K_N(x,y,t)$, i.e.
that $K_N(x,y,0)=\delta(x-y)$, $(\rmi\p_t-h_{Nx})K_N(x,y,t)=0$ and
$(\rmi\p_t-h_{Ny})K_N(x,y,t)=0$ are satisfied. We demonstrate the
explicit proof for the  setup with $u_k(x\to-\infty)\to 0$ omitting
the technically identical considerations for the other cases.

We start by noticing that the intertwiner $L_x$ maps solutions of
the \Sh equation for $h_0$ into solutions of the \Sh equation for
$h_N$ and, hence,  $(\rmi\p_t-h_{Nx})K_N(x,y,t)=0$ is automatically
satisfied.

Next, we consider the initial condition $K_N(x,y;0)=\delta(x-y)$
which should be fulfilled by the r.h.s. of \rf{transNllr}. With
$K_0(x,y,0)=\delta(x-y)$ and $\int_{-\infty}^y
\delta(x-z)u_n(z)dz=\theta(y-x)u_n(x)$ we have from \rf{transNllr}
\bea\label{t4p1}
K_N(x,y;0)&=&(-1)^N\sum_{n=0}^{N-1}(-1)^{n}
\frac{W_{n}(y)}{W(y)}L_{N,0,x}[\theta(y-x)u_n(x)]\,.
\eea
In the Crum-Krein formula (see \rf{operator})
\be\label{t4p2}
L_{N,0,x}[\theta( x-
y)u_n(x)]=\frac{W[u_0,\ldots,u_{N-1},\theta(x-y)u_n(x)]}{W(u_0,\ldots,u_{N-1})}
\ee
we represent the derivatives $\partial_x^m[\theta(y-x)u_n(x)],\
m=0,\ldots, N$ as
\be\label{der}\fl
\partial_x^m[\theta(y-x)u_n(x)]=
\sum_{k=0}^{m-1}C_k^m\theta_x^{(m-k)}(y-x)u_n^{(k)}(x)
+\theta(y-x)u_n^{(m)}(x)\,,
\ee
$\theta_x^{(m-k)}(y-x):=\partial_x^{m-k}\theta(y-x)$. Taking into
account that
\[
\left|\begin{array}{cccc}u_0(x) &  \ldots&u_{N-1}(x)&\theta(y-
x)u_n(x)\\
u'_0(x)&\ldots&u'_{N-1}(x)&\theta(y-x)u'_n(x))\\
\ldots&\ldots&\ldots&\ldots\\
u_0^{(N)}(x)&\ldots&u_{N-1}^{(N)}(x)&\theta(y-x)u_n^{(N)}
\end{array}\right|=0
\]
and making use of the linearity properties of determinants we
reshape \rf{t4p2} as
\bea\label{kn1}\fl\nn
&& L_{N,0,x}[\theta( x-
y)u_n(x)]\\
\fl &=&
\frac 1{W(x)}\left|\begin{array}{cccc}u_0(x) &  \ldots&u_{N-1}(x)&0\\
u'_0(x)&\ldots&u'_{N-1}(x)&-\delta(x-y)u_n(x)\\
\ldots&\ldots&\ldots&\ldots\\
u_0^{(N)}(x)&\ldots&u_{N-1}^{(N)}(x)&\sum_{k=0}^{N-1}C_k^N\theta_x^{(N-k)}(y-
x)u_n^{(k)}(x)
\end{array}\right|.
\eea
Expanding this determinant with regard to the elements of the last
column we find
\bea\label{t4p3}
\fl L_{N,0,x}[\theta( x-
y)u_n(x)]=\frac{(-1)^N}{W(x)}\sum_{m=1}^N(-1)^m
W_{Nm}(x)\sum_{k=0}^{m-1}C_k^m\theta_x^{(m-k)}(y- x)u_n^{(k)}(x)
\eea
where $W_{Nm}(x)$ are corresponding minors. For the verification of
the relation $K_N(x,y;0)=\delta(x-y)$ we use its representation
\be\label{dlt}
\int_{-\infty}^\infty K_N(x,y;0)f(x)dx=f(y)
\ee
where $f(x)$ is a sufficiently smooth test function with compact
support. With \rf{t4p1} and \rf{t4p3} the l.h.s. of \rf{dlt} reads
\be\label{t4p4}\fl
\sum_{n=0}^{N-1}\sum_{m=1}^N\sum_{k=0}^{m-1}(-1)^{n+m}C_k^m
\frac{W_{n}(y)}{W(y)} \int_{-\infty}^\infty dx\,
\theta_x^{(m-k)}(y-x)\,\frac{W_{Nm}(x)f(x)u_n^{(k)}(x)}{W(x)}\,.
\ee
As next step, we use $\theta_x^{(m-k)}(y-x)=-\delta^{(m-k-1)}(x-y)$
and multiple integration by parts\footnote{For the theory of
distributions (generalized functions) see, e.g., \cite{GelShil} (in
particular vol.1 p. 26).} to remove the derivatives from the
$\theta-$functions:
\be\label{t4p5}\fl
\mbox{\rm l.h.s. of \rf{dlt}}=
\sum_{n=0}^{N-1}\sum_{m=1}^N\sum_{k=0}^{m-1}(-1)^{n-k}C_k^m
\frac{W_{n}(y)}{W(y)}\,\p_y^{m-k-1}\left[\frac{W_{Nm}(y)f(y)u_n^{(k)}(y)}{W(y)}\right]\,.
\ee
The relation
\be\label{id}
\frac{1}{W} \sum_{n=0}^{N-1}
(-1)^{n}W_{n}u_{n}^{(j)}=(-1)^N\delta_{j,N-1}\,,\qquad
j=0,\ldots,N-1
\ee
reduces this multiple sum to
\be\label{t4p6}
\mbox{\rm l.h.s. of
\rf{dlt}}=(-1)^N\frac{W_{NN}(y)f(y)}{W(y)}\sum_{k=0}^{N-1}(-1)^k
C_k^{N}
\ee
and because of $W_{NN}(y)=W(y)$ and $\sum_{k=0}^{N-1}(-1)^k
C_k^{N}=(-1)^N$ (cf. 4.2.1.3 in \cite{prudnikov}) the condition
\rf{dlt} is satisfied.

It remains to prove that the \Sh equation
$(\rmi\p_t-h_{Ny})K_N(x,y,t)=0$ is fulfilled too. By explicit
substitution of equation \rf{transNll} we have
\bea\label{t4p7}\fl
&&(\rmi\p_t-h_{Ny})K_N(x,y,t\nn\\
\fl &&=(-1)^N L_{x}\sum_{n=0}^{N-1} (-1)^{n}\frac{W_{n}(y)}{W(y)}
\int_{-\infty}^y
\rmi\p_tK_0(x,z;t)u_{n}(z)dz \nn\\
\fl && +L_{x}\sum_{n=0}^{N-1}
(-1)^{n}\left[\left(\frac{W_{n}(y)}{W(y)}\right)''-
V_N(y)\frac{W_{n}(y)}{W(y)}\right]\int_{-\infty}^y K_0(x,z;t)u_{
n}(z)dz\nn\\
\fl && +L_{x}\sum_{n=0}^{N-1}
(-1)^{n}\left[2\left(\frac{W_{n}(y)}{W(y)}\right)'u_{n}(y)+
\frac{W_{n}(y)}{W(y)}(u_{n}(y)\p_y+u_{n}'(y))\right]K_0(x,y;t)\, .
\eea
First, we note that due to relation \rf{id} and its derivative the
last sum vanishes. Taking further into account that $W_n(y)/W(y)$ is
a solution of the \Sh equation for $h_N$ at energy $E_n$ (cf.
\rf{hNsol}) and replacing $\rmi\p_t K_0(x,z;t)\rightarrow
h_{0z}K_0(x,z;t)$ one reduces equation \rf{t4p7} to
\bea\label{t4p8}\fl
&&(\rmi\p_t-h_{Ny})K_N(x,y,t)\nn\\
\fl &=&(-1)^N L_{x}\sum_{n=0}^{N-1} (-1)^{n}\frac{W_{n}(y)}{W(y)}
\int_{-\infty}^y [(h_{0z}-E_n)K_0(x,z;t)]u_{n}(z)dz\,.
\eea
Integrating by parts and making use of $(h_{0z}-E_n)u_n(z)=0$, the
asymptotical behavior $u(z\to-\infty)\to 0$, $u'(z\to-\infty)\to 0$
and relation \rf{id} one finds that the r.h.s. in \rf{t4p8} vanishes
and, hence, the \Sh equation $(\rmi\p_t-h_{Ny})K_N(x,y,t)=0$ is
fulfilled.
\end{proof}

\subsection{General polynomial supersymmetry}

The three different types of transformations considered above may be
combined in various ways to produce a supersymmetry of more general
type. In general, from the spectrum of the original Hamiltonian
$h_0$ $q$ levels may be removed and $p$ additional levels may be
added, $p+q\le  N$ producing in this way the spectral set of $h_N$.
The inequality would correspond to SUSY transformation chains
between $h_0$ and $h_N$ which contain isospectral transformations.
For further convenience we split the spectra of $h_0$ and $h_N$
according to their transformation related contents as
\bea
\fl \spec(h_0)&=&\{\varepsilon_i,\beta_j,E_k\}+\spec_c(h_0)\,, \ i=1,\ldots, q\,; \ \ j=1,\ldots, N-(p+q+r)\nn\\
\fl \spec(h_N)&=&\{\lambda_l,\beta_j,E_k\}+\spec_c(h_N)\,, \
l=1,\ldots, p\,; \ \ j=1,\ldots, N-(p+q+r)
\eea
where the discrete levels $E_k$ and the continuous spectrum
$\spec_c(h_0)=\spec_c(h_N)$ are not affected by the SYSY
transformations. The set of transformation constants
$\{\a_n\}_{n=0}^{N-1}=\{\varepsilon_i,\lambda_l,\beta_j,\gamma_k\}$
corresponds to $p$ new discrete levels $\lambda_n\in\spec(h_N)$ not
present in $\spec(h_0)$, $q$ levels $\varepsilon_i\in\spec(h_0)$ not
present in $\spec(h_N)$, $N-(p+q+r)$ levels $\beta_j$ present in
both spectra and $r$ constants $\gamma_k$ not coinciding with any
energy level of both Hamiltonians,
$\gamma_k\not\in\spec(h_0)\cup\spec(h_N)$.
Transformations induced
at constants $\a_n=\beta_n,\gamma_n$ are strictly
isospectral.

Summarizing  the previous results the following expression for the
propagator $K_N(x,y,t)$ can be given
\bea\label{trprnp}\nonumber\fl
 K_N(x,y,t)= L_xL_y\sum_{n=0}^{N-1} \left(
\prod_{j=0,j\neq n}^{N-1}\frac{1}{\a_n-\a_j}\right)\int_{-
\infty}^{\infty}K_0(x,z,t)\widetilde{G}_0(z,y,\a_n)dz
\\ 
\hspace{-2em}
+ \sum_{\lambda_n}\phi_{\lambda_n}(x)\phi_{\lambda_n}(y)
\rme^{-\rmi\lambda_nt}
+\sum_{\beta_n}\phi_{\beta_n}(x)\phi_{\beta_n}(y)
\rme^{-\rmi\beta_nt}
\eea
where for $\a_n=\varepsilon_n,\beta_n$
\[
\widetilde{G}_0(z,y,\a_n)= \lim_{E\rightarrow\a_n}
\left[G_0(z,y,E) -\frac{\psi_n(x)\psi_n(y)}{\a_n-E}\right]
\]
 and $\widetilde{G}_0(z,y,\a_n)=G_0(z,y,\a_n)$
otherwise.

\section{Applications\label{applic}}

\subsection{Particle in a box\label{box}}

As first application we consider a free particle in a box, i.e.
the
\Sh equation with $V_0(x)\equiv0$ and Dirichlet BCs at the ends
of the finite interval $(0,1)$. The eigenfunctions of this problem
are the well known $\psi_{n-1}(x)=\sqrt 2\sin(n\pi x)$,
$n=1,2,\ldots$ with energies $E_{n-1}=n^2\pi^2$. The corresponding
propagator reads \cite{Manko}
\[
K_{box0}(x,y,t)=\frac{1}{2}\left[\vartheta_3^{-}(x,y;t)-\vartheta_3^{+}(x,y;t)\right]
\]
with
\[
\vartheta_3^{-}(x,y;t):=
\vartheta_3\left(\frac{x-y}{2}\right|\left.-
\frac{\pi t}{2}\right)\,, \quad
\vartheta_3^{+}(x,y;t):=\vartheta_3\left(\frac
{x+y}{2}\right|\left.-\frac{\pi t}{2}\right)
\]
and  $\vartheta_3(q|\tau)$ denoting the third theta function
\cite{Beitman}.

As SUSY partner problem we choose a model which we derive by
removing the ground state level $E_0$ by a linear (one-step)
SUSY-mapping with $u=\psi_0=\sin \pi x$ as transformation
function\footnote{There exist other types of transformations
leading to regular transformed Sturm-Liouville problems. But the
solutions of the resulting \Sh equations will violate the
Dirichlet BCs \cite{SP} so that a special analysis is needed.}.
This leads to the \Sh equation with  potential $V_1(x)=2\pi^2/\sin^2
(\pi x)$, i.e. a singular Sturm-Liouville problem. The propagator
of this problem can be represented via \eqref{trans1l} as
\be\label{Kbb}\fl
K_{box1}(x,y,t)=-\frac{1}{2\sin\pi
y}L_x\int_{0}^{y}\left[\vartheta_3^{-}(x,z;t)-
\vartheta_3^{+}(x,z;t)\right]\sin(\pi z) dz
\ee
or after explicit substitution of $L_x=-\pi \cot  (\pi
x)+\partial_x $ as
\bea\nonumber\fl
K_{box1}(x,y,t)= \frac{\pi\cot(\pi x)}{2\sin(\pi
y)}\int_{0}^{y}\left[\vartheta_3^{-}(x,z;t)-
\vartheta_3^{+}(x,z;t)\right]\sin(\pi z) dz
\\ \fl
-\frac{\pi}{2\sin(\pi y)}\int_{0}^{y}\left[\vartheta_3^{-}(x,z;t)+
\vartheta_3^{+}(x,z;t)\right]\cos(\pi z) dz
+\frac{1}{2}\left[\vartheta_3^{-}(x,y;t)+
\vartheta_3^{+}(x,y;t)\right].
\eea
Here after using the property
$\p_x\theta_3^{\pm}(x,z,t)=\pm\p_z\theta_3^{\pm}(x,z,t)$
we integrated in \eqref{Kbb} by parts.

\subsection{Harmonic oscillator\label{osci}}

Here we consider the Hamiltonian
\[h_{0}=-\p^2_{xx}+\frac{x^2}{4}\,.\]
Using a two-fold transformation with transformation functions
\[
u_2(x)=(x^2-1)\rme^{-x^2/4}\qquad u_3(x)=x(x^2-3)\rme^{-x^2/4}
\]
corresponding to the second and third excited state
eigenfunctions of $h_0$ we
obtain a perturbed Harmonic oscillator potential \cite{BS}
\be\label{p23}
V^{(2,3)}(x)=\frac{8x^2}{x^4+3}-
\frac{96x^2}{(x^4+3)^2}+\frac{x^2}{4}+2
\ee
which for large $|x|$ behaves like the original harmonic oscillator
potential, but for small $|x|$ shows two shallow minima. For
completeness, we note that the transformation functions $u_2(x)$ and
$u_3(x)$ have nodes, whereas their Wronskian
$W(u_2,u_3)=(x^4+3)\rme^{-x^2/2}$ is nodeless so that the corresponding
second-order SUSY-transformation itself is well defined, but
irreducible.

The propagator for the \Sh equation with Hamiltonian
$h^{(2,3)}=-\p_x^2+V^{(2,3)}(x)$ can be constructed from the well
known propagator
\[
K_{osc}(x,y,t)=
\frac{1}{\sqrt{4\pi \rmi \sin t}}\,
\rme^{\frac{\rmi[(x^2+y^2)\cos t-2xy]}{4\sin t}}\,.
\]
for the $h_0-$\Sh equation (see e.g. \cite{Feynman1}) via relation
\eqref{transNlr}. The occurring integrals
$\int_y^{\infty}K_{osc}(x,z,t)u_n(z)dz$ can be represented as
derivatives with respect to the auxiliary current $J$ of the
generating function
\bea\label{ist}\nonumber\fl
 S(J,x,y,t)&=&
\frac{1}{\sqrt{4\pi \rmi \sin t}}
\int_y^{\infty}\exp\left[\frac{\rmi[(x^2+z^2) \cos t-2xz]}{4\sin
t}-\frac{z^2}{4}+Jz\right]dz
\\
\nonumber \fl &=& \frac{1}{2}
{\exp}{\left(\frac{-2\rmi t-x^2}{4}+
(\rmi J^2\sin t+Jx)\exp({-it})\right)}
\\ \nonumber \fl
&\times & {\left(1+\erf\left[
\frac{-\sqrt{\rmi}\exp({-\frac{\rmi t}{2}})
\left(2J\sin t+\rmi(y\exp({\rmi t})-x)\right)}{2\sqrt{\sin t}}
\right]\right)}\,.
\eea
As a result, we obtain
\bea\nonumber
K^{(2,3)}(x,y,t)
\\ \nonumber \fl
=\rme^{y^2/2}L_x\left( \frac{y(y^2-3)}{y^4+3} \left[\frac{\p^2S(J)}{\p
J^2}-S(J)\right]-\frac{y^2- 1}{y^4+3}\left[\frac{\p^3S(J)}{\p J^3}-
3\frac{\p S(J)}{\p J}\right]\right)_{J=0}
\eea
in terms of obvious abbreviations.

The technique can be straight forwardly generalized to second-order
SUSY-transformations built on any eigenfunction pair
\[
u_k(x)=p_k(x)\rme^{-x^2/4}\qquad
u_{k+1}(x)=p_{k+1}(x)\rme^{-x^2/4}
\]
of the Hamiltonian $h_0$. The corresponding generalized potentials
$V^{(k,k+1)}(x)$ (see \cite{BS}) read
\be\label{pkk1}
V^{(k,k+1)}(x)=-
2\frac{Q_k''(x)}{Q_k(x)}+
2\left[\frac{Q_k'(x)}{Q_k(x)}\right]^2+\frac{x^2}{4}+4
\ee
where
\[ \fl
Q_k(x):=p_{k+1}^2(x)-p_k(x)p_{k+2}(x)\qquad
W(u_k,u_{k+1})=-Q_k(x)\exp(-x^2/2)
\]
are built from  the re-scaled Hermite polynomials
$p_k(x)=2^{-k/2}H_k(x/\sqrt2)$.

With the help of \eqref{transNl} the propagator is obtained as
\be\nonumber\fl
K^{(k,k+1)}(x,y,t)
=\rme^{y^2/2}L_x\left( \frac{p_{k+1}(y)}{Q_k(y)}
\left[p_{k}\left(\p_J\right)S(J)\right]-\frac{p_{k}(y)}{Q _k(y)}
\left[p_{k+1}\left(\p_J\right)S(J)\right]\right)_{J=0}
\ee
with $p_{k}\left(\p_J\right)$ denoting the $k$th-order differential
operator obtained from $p_k(x)$ by replacing $x^n\rightarrow
\frac{\p^n}{\p J^n}$.

Finally we note that the potentials \eqref{pkk1} behave
asymptotically like $x^2/4$ for $|x|\to \infty$ and have $k$ shallow
minima at their bottom.

\subsection{Transparent potentials\label{transparent-pot}}

Here we  apply our propagator calculation method to transparent
potentials which are SUSY partners of the zero potential in models
defined over the whole real axis. We note that the propagator for
a one-level transparent potential was previously calculated by
Jauslin \cite{jauslin} for the Fokker-Planck equation. The
propagator for a two-level potential can be found in \cite{SP}
where the general form of the propagator for an $N$-level
potential has been given as a conjecture. Here we will prove this
conjecture.

We construct the $N$-level transparent potential from the zero
potential with the help of the following $N$ solutions of the \Sh
equation with $V_0(x)\equiv0$ (see e.g. \cite{BS})
\bea\label{fpfs} &&
u_{2j-1}(x)=\cosh(a_{2j-1}x+b_{2j-1})\\
&&\nonumber u_{2j}(x)=\sinh(a_{2j}x+b_{2j})\qquad
j=1,2,\ldots,[(N+1)/2]
\eea
as transformation functions. Here, $a_i,b_i\in \RR$\,,
$a_i>a_{i+1}>0$ is assumed and $[(N+1)/2]$ denotes the integer
part of $(N+1)/2$\,.

The factorization constants $\a_j=-a_j^2$ define the positions of
the discrete levels (point spectrum)  $E_j=\a_j<0$ of
$h_N=-\partial_x^2+V_N(x)$, whereas the continuous part of the
spectrum of $h_N$ fills the whole real axis.
The eigenfunctions of
the discrete levels normalized to unity have the form
\cite{Sukumar1}
\begin{equation}\label{ortef}\fl
\varphi_n(x)= \left[\frac{a_n}{2} \prod_{j=1,j\neq
n}^{N}|a_n^2-a_j^2|\right]^{1/2}
\frac{W(u_1,u_2,\ldots,u_{n-1},u_{n+1},\ldots,u_N)}%
{W(u_1,u_2,\ldots,u_N)}\,.
\end{equation}
The propagator $K_0$ and the Green function $G_0$ for the free
particle are well-known \cite{Grosche}
\bea\label{K0free}
K_0(x,y;t)=\frac{1}{\sqrt{4\pi \rmi t}}\,
 \rme^{ \frac{\rmi(x-y)^2}{4t}}
\\\nonumber
G_0(x,y,E)=\frac{\rmi}{2\k}\rme^{\rmi\k |x-y|}\qquad
\mbox{Im}\k>0\quad\quad E=\k^2
\eea
so that according to (\ref{trprn}) the propagator of the
transformed system can be calculated as
\bea\label{k1}\nonumber && \fl
K_N(x,y,t)=\frac{L_xL_y}{\sqrt{4\pi \rmi t}} \sum_{n=1}^N \left(
\prod_{j=1,j\neq n}^N\frac{1}{\a_n-\a_j} \right)
\int_{-\infty}^{\infty}
\exp\left(\frac{\rmi(x-z)^2}{4t}-a_n|z-y|\right)\frac{dz}{2a_n}
\\
&&+\sum_{n=1}^{N}\varphi_n(x)\varphi_n(y)\rme^{-\rmi\a_nt}
=:K_{cN}(x,y,t)+K_{dN}(x,y,t)\,.
\eea
In the last line we separated contributions from the continuous
spectrum, $K_{cN}(x,y,t)$, from contributions from the discrete
spectrum, $K_{dN}(x,y,t)$.

The integral in $K_{cN}(x,y,t)$ can easily be calculated since the
primitive of the integrand is well known (see e.g. integral 1.3.3.17
of Ref \cite{prudnikov}). Using the well studied convergency
conditions of the error-function erfc from \cite{abramowitz} we find
\bea\nonumber\fl
I(a,x,y,t):= \frac{1}{\sqrt{4\pi \rmi t}} \int_{-\infty}^{\infty}
\exp\left(\frac{\rmi(x-z)^2}{4t}-a|z-y|\right)\frac{dz}{2a}\\
\fl = \frac{\rme^{\rmi a^2t}}{4a}\left[
\rme^{a(x-y)}
\mbox{erfc}\left(a\sqrt{\rmi t}+\frac{x-y}{2\sqrt{\rmi t}}
\right) +
\rme^{a(y-x)}
\mbox{erfc}\left(a\sqrt{\rmi t}-\frac{x-y}{2\sqrt{\rmi t}}
\right)
\right]\\
\fl a>0\qquad t>0\nonumber
\eea
and with it
\be\label{kc}
K_{cN}(x,y;t) = L_xL_y\sum_{n=1}^N\left(\prod_{m=1,m \neq n}^N
\frac{1}{\alpha_n-\alpha_m}\right) I(a_n,x,y,t)\,.
\ee

Introducing the notation
\[
\erf_{\pm}(a):=
\erf\left(a\sqrt{it}\mp\frac{x-y}{2\sqrt{\rmi t}}\right)
\]
and abbreviating the Wronskian of the $x-$dependent
transformation functions $u_1$,\ldots, $u_N$ as $W(x)$ we
formulate the final expression for the propagator as
\begin{Th}\label{th4}
The propagator for an $N$-level transparent potential has the form
\bea\label{nspp}\nonumber
K_N(x,y;t)= \frac{1}{\sqrt{4\pi \rmi t}}\, \rme^{ \frac{\rmi(x-y)^2}{4t}}
\\
\fl +\sum_{n=1}^{N}\left(\frac{a_n}{4} \prod_{j=1,j\neq
n}^{N}|a_n^2-a_j^2|\right)
\frac{W_{n}(x)W_{n}(y)}{W(x)W(y)}\,\rme^{\rmi a_n^2t} \left[\erf_{+}(a_n)
+\erf_{-}(a_n)\right]\,.
\eea
\end{Th}
For the proof of this theorem we need the additional
\begin{Lm} \label{Lm3}
Let $\{\a_i\}_{i=1}^N$  be a set of $N$ non-coinciding complex
numbers $a_i\neq a_{j\neq i}\in \CC$ and\footnote{We use the
standard notation $\ZZ^+=\{0,1,2,\ldots\}$ for the natural numbers
with zero included.} $n\in \ZZ^+$. Then the following identity
holds:
\be\label{s1}
\sum_{i=1}^N \a_i^n\left(\prod_{j=1,j\neq
i}^N\frac{1}{\a_i-\a_j}\right)=\delta_{n,N-1}\,.
\ee
\end{Lm}
\begin{proof}
Consider the auxiliary function
\[
f(z)=\frac{z^n}{(z-\a_1)(z-\a_2)\ldots(z-\a_N)}
\]
which is analytic in any finite part of the complex $z$-plane except
for $N$ simple poles $\a_1,\ldots,\a_N$. From the residue theorem
follows
\[
\sum_{i=1}^N \res f(\a_i)=\sum_{i=1}^N
\a_i^n\left(\prod_{j=1,j\neq i}^N\frac{1}{\a_i-\a_j}\right)
=-\res f(\infty)
\]
what with the residue at infinity yields (\ref{s1}).
\end{proof}
As the next step we prove Theorem \ref{th4}.
\begin{proof}
Without loss of generality we may set $b_j=0$. We start with the
propagator component $K_{cN}(x,y,t)$ in \eqref{kc} related to the
continuous spectrum. First we note that the function $I$ in
\eqref{kc} depends only via the difference $x-y$ on the spatial
coordinates so that the action of $\p_y$ in $L_y$ can be replaced by
$\p_y^n\rightarrow(-1)^n\p_x^n$. Hence, the composition of the two
$N$th-order transformation operators $L_xL_y$ acts as an effective
$2N$th-order differential operator in $x$
\be\label{L2x}
L_xL_y=R_0+R_1\p_x+\ldots+R_{2N}\p_x^{2N}
\ee
with coefficient functions $R_n(x,y)$.  Accordingly, \eqref{kc}
takes the form
\be\label{kcc}
\fl K_{cN}(x,y;t) =\sum_{n=1}^N\left(\prod_{m=1,m \neq
n}^N\frac{1}{\alpha_n-
\alpha_m}\right)[R_0+R_1\p_x+\ldots+R_{2N}\p_x^{2N}] J_n
\ee where
$J_n=I\left(a_n,x,y,t\right)$.
From the explicit structure of the
first derivatives of $J_n$
\bea
\fl\nonumber \frac{\p J_n}{\p x}= \frac{1 }{4}\left[-\rme^{-a_n(x-y)}
\erfc_{+}(a_n) +\rme^{a_n(x-y)} \erfc_{-}(a_n) \right]\rme^{\rmi a_n^2t}
\\\nonumber
\fl \frac{\p^2 J_n}{\p x^2}=\frac{-1}{\sqrt{4\pi \rmi t}}\,
\rme^{\frac{\rmi(x-y)^2}{4t}}+a_n^2I(a_n)
\\
 \fl \nonumber
\frac{\p^3 J_n}{\p x^3}=\frac{-1}{ 2t\sqrt{4\pi \rmi t}}
\,(x-y)\,\rme^{\frac{\rmi(x-y)^2}{4t}}
\\ \fl
 +\frac{a_n^2
}{4}\left[-\rme^{-a_n(x-y)} \erfc_{+}(a_n) +\rme^{a_n(x-y)}
\erfc_{-}(a_n) \right]\rme^{\rmi a_n^2 t}
\eea
we find (by induction) the general expression of an even-order
derivative
\bea \nonumber
\frac{\p^{2m}I}{\p x^{2m}}&=&
\sum_{k=0}^{m- 1}a_n^{2k}A_{km}(x,y,t)
\rme^{\frac{\rmi(x-y)^2}{4t}}\\
&+&\frac{a_n^{m-1}}{4}
\rme^{\rmi a_n^2 t}\left[\rme^{-a_n(x-y)}\erfc_{+}(a_n)+\rme^{a_n(x-y)}
\erfc_{-}(a_n)\right]
\eea
and of an odd-order derivative
\bea\nonumber\hspace{-1em}
\frac{\p^{2m-1}I}{\p x^{2m-1}}&=&
\sum_{k=0}^{m- 2}a_n^{2k}A_{km}(x,y,t)
\rme^{\frac{\rmi(x-y)^2}{4t}}\\
 &-&\frac{a_n^{m-1}}{4}
\rme^{\rmi a_n^2 t}\left[\rme^{-a_n(x-y)}\erfc_{+}(a_n)-\rme^{a_n(x-y)}
\erfc_{-}(a_n)\right]\,.
\eea
It holds
\[
A_{km}(x,y,t)\equiv 0\qquad \Longleftrightarrow\qquad
\left\{\begin{array}{lcc}
  m=2l & \cap &  k>l-1\\
  m=2l+1 &\cap & \hspace{.5em} k>l-1\,. \\
\end{array}\right.
\]
Subsequently we use the abbreviation
\be \label{In}\fl
I_m(a_n)=\frac{1}{4 a_n}\,
\rme^{\rmi a_n^2t}
\left[(-1)^m\rme^{-a_n(x-y)}\erfc_{+}(a_n)+\rme^{a_n(x-y)}
\erfc_{-}(a_n)\right]
\ee
and we will need the explicit form of the $a_n^{2(N-1)}$ and
$a_n^{2N}$ terms in the highest-order derivative
\be\label{hd}
\frac{\p^{2N}J_n}{\p x^{2N}}=
\ldots -a_n^{2(N-1)}\frac{1}{\sqrt{4\pi \rmi t}}\,
\rme^{\frac{\rmi(x-y)^2}{4t}}
+a_n^{2N}I_{2N}\,.
\ee
The complete propagator component $K_{cN}(x,y;t)$ in (\ref{kcc})
can now be rewritten as
\bea\nonumber
K_{cN}(x,y;t) =\sum_{n=1}^N\left(\prod_{m=1,m \neq
n}^N\frac{1}{\alpha_n-\alpha_m}\right)R_0I_0(a_n)
\\ \nonumber
+ \sum_{n=1}^N \left(\prod_{m=1,m \neq
n}^N\frac{1}{\alpha_n-\alpha_m}\right) a_nR_1I_1(a_n)
\\ \nonumber
+ \sum_{n=1}^N \left(\prod_{m=1,m \neq
n}^N\frac{1}{\alpha_n-\alpha_m}\right)
R_{2}\left[-\frac{1}{\sqrt{4\pi \rmi t}}\,
\rme^{\frac{\rmi(x-y)^2}{4t}}+a_n^2I_2(a_n)\right]+\ldots
\\ \nonumber \fl
+\sum_{n=1}^N
\left(
\prod_{m=1,m \neq n}^N\frac{1}{\alpha_n-\alpha_m}
\right)
R_{2N}
\left[\sum_{k=0}^{N-1}a_n^{2k}A_{k,2N}(x,y,t)\,
\rme^{\frac{\rmi(x-y)^2}{4t}}+a_n^{2N}I_{2N}(a_n)
\right] .
\nonumber
\eea
Comparison with Lemma \ref{Lm3} shows that due to
$a_n^{2k}=(-\a_n)^k$ only a single term containing
$\exp{({\rmi(x-y)^2}/{4t}})$ does not vanish.
It is the $k=N-1$ term
in the very last sum which with $R_{2N}=(-1)^N$ and
$A_{N-1,2N}(x,y,t)=-(4\pi \rmi t)^{-1/2}$ yields exactly the
propagator
\eqref{K0free}
of the free particle.
All other terms containing
$\exp({{\rmi(x-y)^2}/{4t}})$, after interchanging the sums
$\sum_{n=1}^N\ldots $ and $\sum_{k=0}^{N-1}\ldots $,
 cancel out due to
Lemma \ref{Lm3}.
Thus, the above formula reduces to
\bea\label{ksol}\fl
K_{cN}(x,y;t) =\frac{1}{\sqrt{4\pi \rmi t}}\rme^{ \frac{\rmi(x- y)^2}{4t}}+
\sum_{n=1}^N\left(\prod_{m=1,m \neq n}^N\frac{1}{\alpha_n-
\alpha_m}\right)
\\
\nonumber \fl
\times\left[R_0I_0(a_n)+R_1a_nI_1(a_n)+R_{2}a_n^2I_2(a_n)+
R_{3}a_n^3I_3(a_n)+\ldots+R_{2N}a_n^{2N}I_{2N}(a_n)\right].
\eea
Expressing $I_m(a_n)$ via \eqref{In} in terms of $\mbox{erfc}_+$
and $\mbox{erfc}_-$ one finds contributions
\[\fl
\frac{ \erfc_{+}(a_n)}{4a_n}
\rme^{\rmi a_n^2t}(R_0-a_nR_1+a_n^2R_{2}-
a_n^3R_{3}+\ldots+a_n^{2N}R_{2N})\rme^{-a_n(x-y)}
\]
\[\fl
\frac{ \erfc_{-}(a_n) }{4a_n}
\rme^{\rmi a_n^2 t}(R_0+a_nR_1+a_n^2R_{2}+
a_n^3R_{3}+\ldots+a_n^{2N}R_{2N})\rme^{a_n(x-y)}
\]
which can be represented as
\[\fl
\frac{ \erfc_{+}(a_n)}{4a_n}
\rme^{\rmi a_n^2 t}(R_0+R_1\p_x+R_{2}\p_x^2+
R_{3}\p_x^3+\ldots+R_{2N}\p_x^{2N})\rme^{-a_n(x-y)}
\]
\[\fl
\frac{\erfc_{-}(a_n)}{4a_n}
\rme^{\rmi a_n^2t}(R_0+R_1\p_x+R_{2}\p_x^2+
R_{3}\p_x^3+\ldots+R_{2N}\p_x^{2N})\rme^{a_n(x-y)}\,.
\]
Comparison with \eqref{L2x} shows that the sums  yield simply
$L_xL_y \exp[\pm a_n (x-y)]$. Recalling furthermore that the
transformation operators $L_x$ and $L_y$ are given by
\eqref{operator} and that the transformation functions have the form
\eqref{fpfs} we arrive after some algebra at\footnote{We note that
equation \rf{s0} is compatible with \eqref{ortef}. The function
$\exp(-a_nx)$ is a solution of the initial \Sh equation related to
one of the factorization constants and it is linearly independent
from the corresponding factorization solution. Therefore
$L_x\exp(-a_nx)$ up to a constant factor is one of the bound state
functions of $h_N$ given in \eqref{ortef}.}
\be\label{s0}
L_x\rme^{\pm a_nx}= (\mp1)^n(-1)^{N+n-1}a_n\prod_{j=1,j\neq n}^{N}
(a_j^2- a_n^2)\frac{W_{n}(x)}{W(x)}
\ee
and, hence, at
\be\label{sl}
L_xL_y\rme^{\pm a_n(x-y)}= (-1)^na_n^2\prod_{j=1,j\neq n}^{N} (a_n^2-
a_j^2)^2\frac{W_{n}(x)W_{n}(y)}{W(x)W(y)}\,.
\ee
Substituting \eqref{sl} into \eqref{ksol} we obtain
\bea\label{ksol1}\fl\nonumber
 K_{cN}(x,y;t) =
\frac{1}{\sqrt{4\pi \rmi t}}\,\rme^{ \frac{\rmi(x-y)^2}{4t}} +
\frac{1}{4}
\sum_{n=1}^N {\rm{e}}^{ia_n^2 t}a_n
\left(\prod_{m=1,m \neq n}^N
\frac{(-1)^n(a_n^2- a_m^2)^2}{\alpha_n-\alpha_m}\right)
\\\hspace{-2em}
\times\frac{W_{n}(x)W_{n}(y)}{W(x)W(y)}\left[\erfc_{+}(a_n)
 +\erfc_{-}(a_n)\right].
\eea
A further simplification can be achieved by recalling that
$\a_n=-a_n^2$ and that the parameters $a_i$ are ordered as
$a_1<a_2<\ldots<a_N$. Setting $(a_m^2-a_n^2)=-(a_n^2-a_m^2)$ in the
denominator for $m=1,\ldots,n-1$ gives an additional sign factor
$(-1)^{n-1}$. The propagator sum $K_N=K_{cN}+K_{dN}$ resulting from
continuous and discrete spectral components reads then
\bea\label{ksol2}\fl\nonumber
K_{N}(x,y;t) = \frac{1}{\sqrt{4\pi \rmi t}}\ \rme^{ \frac{\rmi(x-y)^2}{4t}}
-\frac{1}{4}
\sum_{n=1}^N {\rm{e}}^{\rmi a_n^2 t}a_n
\left(\prod_{m=1,m\neq n}^N|a_n^2- a_m^2|\right)
\\ \nonumber\hspace{-2em}
\times\frac{W_{n}(x)W_{n}(y)}{W(x)W(y)}\left[\erfc_{+}(a_n)
 +\erfc_{-}(a_n)
\right]\\\hspace{-2em}
+\frac{1}{2} \sum_{n=1}^N {\rm{e}}^{\rmi a_n^2 t}a_n
\left(\prod_{m=1,m\neq n}^N|a_n^2- a_m^2|\right)
\frac{W_{n}(x)W_{n}(y)}{W(x)W(y)} \,.
\eea
and via the relation $\mbox{erfc}(z)=1-\mbox{erf}(z)$ it leads to
the expression in the theorem.
\end{proof}

\section{Conclusion\label{conclu}}
A careful study of propagators for SUSY partner Hamiltonians is
given. Assuming the partner Hamiltonians linked by polynomial
supersymmetry of a general type we derived user friendly
expressions interrelating the corresponding associated
propagators. Since the propagators may be also defined in terms of
continual integrals the results should be useful in exploring new
classes of continual integrals. We applied our general technique
to derive propagators for transparent potentials and for a family
of SUSY partner potentials of the Harmonic Oscillator. Further
applications can be expected in time-evolution scenarios for
particles placed in corresponding potential fields, in stochastic
processes described by Focker-Planck equations (see e.g.
\cite{jauslin}), in statistical physics and possibly in dynamo
processes of magnetohydrodynamics \cite{GSS}.

\section*{Acknowledgments}
The work has been partially supported by the grants RFBR-06-02-16719
(B.F.S. and A.M.P), SS-5103.2006.2 (B.F.S.), by the Russian Dynasty
Foundation (A.M.P.) and by the German Research Foundation DFG, grant
GE 682/12-3 (U.G.).

\appendix
\section{\label{ap1}}

Here we derive a representation of the unphysical auxiliary
solutions $u_{N,n}$, $n=0,\ldots,N-1$ of the $h_N-$\Sh equation in
terms of Wronskian fractions. We start by recalling that the
functions $u_{N,n}$ are defined by acting with an $N$th-order
(polynomial) SUSY-transformation operator $L_{N,0}$ on the
unphysical solutions $\widetilde u_{0,n}$ of the \Sh equation with
Hamiltonian $h_0$ at energies $E_n=\a_n$. The functions are linearly
independent from their physical counterparts $u_{0,n}$ and
normalized as $W(u_{0,n},\widetilde u_{0,n})=1$. The operator
$L_{N,0}$ itself is constructed from the $u_{0,n}$, $n=0,\ldots,N-1$
as transformation functions.

Below we show by induction that the unphysical solutions
$u_{N,0}$ defined in \rf{LN0u} as
\[
u_{N,n}=L_{N,0}\widetilde{u}_{0,n}=L_{N,0}u_{0,n}\int_{x_0}
^x\frac{dy}{u_{0,n}^2(y)}
\]
have the following representation in terms of Wronskian fractions
\bea\fl\label{norm}
u_{N,n}=C_{N,n}
\frac{W_n(u_{0,0},\ldots,u_{0,N-1})}{W(u_{0,0},\ldots,u_{0,N-1})}\\
\fl C_{N,n}:=(E_{N-1}-E_n)(E_{N-2}-E_n)
\ldots(E_{n+1}-E_n)\qquad n=0,\ldots,N-2\nn\\
\fl u_{N,N-1}=
\frac{W_{N-1}(u_{0,0},\ldots,u_{0,N-1})}{W(u_{0,0},\ldots,u_{0,N-1})}
=\frac{W(u_{0,0},\ldots,u_{0,N-2})}{W(u_{0,0},\ldots,u_{0,N-1})}
\,.\label{A2}
\eea

For $N=1$ the first order transformation operator $L_{1,0}$ is
constructed with the help of $u_{0,0}$. Therefore applying
\eqref{L1} with $f=\widetilde{u}_{0,0}$ yields
\[
u_{1,0}=L_{1,0}\widetilde{u}_{0,0}=
\frac{1}{u_{0,0}}
\]
which obviously agrees with the statement.
Applying \eqref{L1} with $f={u}_{0,1}$ we obtain
\be\label{u11}
u_{1,1}=L_{1,0}u_{0,1}=
\frac{W(u_{0,0},u_{0,1})}{u_{0,0}}\,.
\ee

In order to prove the statement for $N=2$ we build the linear
SUSY-operator $L_{2,1}$ from $u_{1,1}$ and act with it on the
unphysical solutions  $\widetilde{u}_{1,1}$ and $u_{1,0}$. As
first result we obtain
\[
u_{2,1}=L_{2,1}\widetilde u_{1,1}=\frac{1}{u_{1,1}}=
\frac{u_{0,0}}{W(u_{0,0},u_{0,1})}
\]
where \eqref{L1} and \eqref{u11} have been used. For the second
function we find
\bea\fl\nonumber
u_{2,0}=L_{2,1}u_{1,0}=L_{2,1}\frac{1}{u_{0,0}}=\frac{1}{u_{0,0}u_
{1,1}}L_{1,0}^{+}u_{1,1}
=\frac{1}{u_{0,0}u_{1,1}}L_{1,0}^{+}L_{1,0}u_{0,1}
\\\hspace{-.5em}
=(E_1-E_0)\frac{1}{u_{0,0}u_{1,1}}u_{0,1}=
(E_1-E_0)\frac{u_{0,1}}{W(u_{0,0},u_{0,1})}\,.
\eea
Here the operator $L_{2,1}=-u_{1,1x}/u_{1,1}+\p_x$  has been
replaced by $L^+_{1,0}=-u_{0,0x}/u_{0,0}-\p_x$ with the help of
$(1/u_{0,0})'=-u'_{0,0}/u^2_{0,0}$. Moreover, we have used
\eqref{u11}, the factorization rule $L_{1,0}^{+}L_{1,0}=h_0-E_0$
and the \Sh equation $h_0u_{0,1}=E_1u_{0,1}$.

Assuming finally that the representations \eqref{norm} and
\eqref{A2} are valid for $u_{N,n}$ and $u_{N,N-1}$ we prove them
for $u_{N+1,n}$ and $u_{N+1,N}$.

We start with $u_{N,N-1}\mapsto u_{N+1,N}$. To go from $h_N$ to
$h_{N+1}$ only the linear (one-step) SUSY transformation $L_{N+1,N}$
is required. Applying \rf{L1} and combining it with  the
normalization condition $W(u_{N,N},\wt u_{N,N})=1$ and the
Crum-Krein formula \rf{operator} [or \rf{uNN}] we find the
equivalence chain
\[\fl
u_{N+1,N}= L_{N+1,N}\wt u_{N,N}=\frac{W(u_{N,N},\wt
u_{N,N})}{u_{N,N}}= \frac{1}{u_{N,N}}=
\frac{W(u_{0,0},\ldots,u_{0,N-1})}{W(u_{0,0},\ldots,u_{0,N})}\,.
\]
Comparison with \rf{A2} shows that the proof is done.

The proof of the induction $u_{N,n}\mapsto u_{N+1,n}$ is less
obvious. In section \ref{4.2} the linear intertwiners $L_{m+1,m}$
have been build strictly incrementally from the corresponding ground
state eigenfunctions $u_{m,m}$ (of the Hamiltonians $h_m$) as
transformation functions. Here, we need a more general
non-incremental construction scheme. In order to facilitate it, we
first introduce a very detailed notation for general polynomial
intertwiners $L_{k,0}$ indicating explicitly the energy levels of
the transformation functions from which they are built. Based on the
Crum-Krein formula \rf{operator} we set
\be\label{L-def-expl}
L^{(a_1,a_2,\ldots,a_k)}_{k,0}f:=
\frac{W(u_{0,a_1},u_{0,a_2},\ldots,u_{0,a_k},f)}%
{W(u_{0,a_1},u_{0,a_2},\ldots,u_{0,a_k})}\qquad
a_i\neq a_{j\neq i}
\ee
with $a_i\in\ZZ^+$ being any energy level numbers of the Hamiltonian
$h_0$. The determinant structure of \rf{L-def-expl} immediately
implies the generalized kernel property
\be
L_{k,0}^{(a_1,\ldots,n,\ldots, a_k)}u_{0,n}=0
\ee
and the invariance of the operator $L^{(a_1,a_2,\ldots,a_k)}_{k,0}$
with regard to permutations $(a_1,a_2,\ldots,a_k)\mapsto
\sigma(a_1,a_2,\ldots,a_k)$
\be
L^{\sigma(a_1,a_2,\ldots,a_k)}_{k,0}=L^{(a_1,a_2,\ldots,a_k)}_{k,0}\,.\label{permute}
\ee
Recalling that \rf{L-def-expl} can be built from a chain
$L_{k,k-1}^{(a_k)}L_{k-1,k-2}^{(a_{k-1})}\cdots L_{1,0}^{(a_1)}$ of
linear intertwiners $L_{m+1,m}^{(a_m)}$ we conclude that the
ordering with regard to energy levels is inessential in the
transformation chain and that we can split it into sub-chains with
any permuted combination of levels
\bea
\sigma(a_1,\ldots,a_k)=(b_1,\ldots,b_B,c_1,\ldots,c_C)\,,\quad
B+C=k\nn\\
L_{k,0}^{(a_1,\ldots,a_{k})}=L_{B+C,B}^{(c_1,\ldots,c_C)}L_{B,0}^{(b_1,\ldots,b_B)}=L_{C+B,C}^{(b_1,\ldots,b_B)}L_{C,0}^{(c_1,\ldots,c_C)}\,.
\label{commute}
\eea
Apart from this full commutativity of the transformations we note
their associativity
\be
\fl L_{m+3,m+2}^{(a_3)}\left(L_{m+2,m+1}^{(a_2)}\,
L_{m+1,m}^{(a_{1})}\right) =\left(L_{m+3,m+2}^{(a_3)}\,
L_{m+2,m+1}^{(a_2)}\right)L_{m+1,m}^{(a_{1})}\label{associate}\,.
\ee
Commutativity and associativity can be used to re-arrange a sequence
of transformations in any required order.

In accordance with the intertwiners  $L$ we denote
eigenfunctions as
\be\label{u-def-expl}
u_{k,n}^{(a_1,a_2,\ldots,a_k)}:=L^{(a_1,a_2,\ldots,a_k)}_{k,0}u_{0,n}\,,\qquad
a_i\neq n\,,\ i=1,\ldots, k\,.
\ee
Another ingredient that we need is the representation
\bea\fl
\frac{W(u_{0,0},\ldots,u_{0,N-1})}{W_n(u_{0,0},\ldots,u_{0,N-1})}
&=&(-1)^{N-1-n}
\frac{W(u_{0,0},\ldots,u_{0,n-1},u_{0,n+1},\ldots,u_{0,N-1},u_{0,n})}%
{W_n(u_{0,0},\ldots,u_{0,N-1})}\nn\\
\fl&=&(-1)^{N-1-n}L^{(0,\ldots,n-1,n+1,\ldots,N-1)}_{N-1,0}u_{0,n}\nn\\
\fl&=&(-1)^{N-1-n}u^{(0,\ldots,n-1,n+1,\ldots,N-1)}_{N-1,n}=:v_{N-1,n}
\label{Wn-frac}
\eea
which immediately follows from the Crum-Krein formula
\rf{L-def-expl} and definition \rf{u-def-expl}.  From the physical
solution $v_{N-1,n}$ we build the operators%
\footnote{For $C\neq
0$ holds $L^{(n)}_{N,N-1}[Cv_{N,n}]=L^{(n)}_{N,N-1}[v_{N,n}]$ so
that the sign factor $(-1)^{N-1-n}$ plays no role in the operator
$L^{(n)}_{N,N-1}$ itself.
}
\bea
L_{N,N-1}^{(n)}&:=&L_{N,N-1}^{(n)}[v_{N-1,n}]=
-\frac{v_{N-1,n,x}}{v_{N-1,n}}+
\p_x\nn\\
L_{N,N-1}^{(n)+}&:=&L_{N,N-1}^{(n)+}[v_{N-1,n}]=
-\frac{v_{N-1,n,x}}{v_{N-1,n}}-
\p_x\label{Lnv}
\eea
and the corresponding Hamiltonian
$h_{N-1}^{(0,\ldots,n-1,n+1,\ldots,N-1)}$ with
\be
L_{N,N-1}^{(n)+}L_{N,N-1}^{(n)}=h_{N-1}^{(0,\ldots,n-1,n+1,\ldots,N-1)}-E_n\,.
\ee

We start the proof of the induction $u_{N,n}\mapsto u_{N+1,n}$
with the following transformations
\bea
u_{N+1,n}&=&L^{(N)}_{N+1,N}\,u_{N,n}=
C_{N,n}\,L^{(N)}_{N+1,N}\,\frac{W_n(u_{0,0},\ldots,u_{0,N-1})}%
{W(u_{0,0},\ldots,u_{0,N-1})}\nn\\
&=&C_{N,n}\,L^{(N)}_{N+1,N}\,v_{N-1,n}^{-1}\nn\\
&=&C_{N,n}\left[-\frac{u_{N,N,x}}{u_{N,N}v_{N-1,n}}-
\frac{v_{N-1,n,x}}{v_{N-1,n}^2}\right]\nn\\
&=&C_{N,n}\frac1{u_{N,N}v_{N-1,n}}\,L_{N,N-1}^{(n)+}\,u_{N,N}\label{uNnproof1}\,.
\eea
In order to obtain the operator product
$L_{N,N-1}^{(n)+}L_{N,N-1}^{(n)}$ we use \rf{uNN}, the composition
rule \rf{commute} and the Crum-Krein formula \rf{L-def-expl}
\bea\label{uNnproof2}
u_{N,N}&=&L_{N,0}^{(0,1,\ldots,N-1)}u_{0,N}=
\frac{W(u_{0,0},\ldots,u_{0,N})}{W(u_{0,0},\ldots,u_{0,N-1})}\nn\\
&=&L_{N,N-1}^{(n)}L_{N-1,0}^{(0,1,\ldots,n-1,n+1,\ldots,N-1)}u_{0,N}\nn\\
&=&L_{N,N-1}^{(n)}u_{N-1,N}^{(0,1,\ldots,n-1,n+1,\ldots,N-1)}
\eea
where
\bea
u_{N-1,N}^{(0,1,\ldots,n-1,n+1,\ldots,N-1)}
&=&L_{N-1,0}^{(0,1,\ldots,n-1,n+1,\ldots,N-1)}u_{0,N}\nn\\
&=&\frac{W_n(u_{0,0},\ldots,u_{0,N})}{W_n(u_{0,0},\ldots,u_{0,N-1})}
\label{uN-1N}\,.
\eea
This gives
\bea
u_{N+1,n}&=&C_{N,n}\frac1{u_{N,N}v_{N-1,n}}\,L_{N,N-1}^{(n)+}\,u_{N,N}\nn\\
&=&C_{N,n}\frac1{u_{N,N}v_{N-1,n}}\,
L_{N,N-1}^{(n)+}L_{N,N-1}^{(n)}u_{N-1,N}^{(0,1,\ldots,n-1,n+1,\ldots,N-1)}\nn\\
&=&C_{N,n}\frac1{u_{N,N}v_{N-1,n}}
(E_N-E_n)u_{N-1,N}^{(0,1,\ldots,n-1,n+1,\ldots,N-1)}\nn\\
&=&C_{N+1,n}\frac{W_n(u_{0,0},\ldots,u_{0,N})}{W(u_{0,0},\ldots,u_{0,N})}
\label{uNnproof3}
\eea
where the last line has been obtained by expressing
$u_{N-1,N}^{(0,1,\ldots,n-1,n+1,\ldots,N-1)}$, $u_{N,N}$ and
$v_{N-1,n}$  via \rf{uN-1N}, \rf{uNN} and \rf{Wn-frac}  in terms of
their Wronskian fractions. With \rf{uNnproof3} the proof is
complete.

\end{document}